\documentclass[a4paper,10pt]{article}

\pdfoutput=1

\usepackage{jheppub}

\usepackage{graphicx}   
\usepackage{dcolumn}    
\usepackage{bm}         
\usepackage{verbatim}
\usepackage{environ}

\usepackage[english]{babel}
\usepackage{xcolor}
\usepackage{graphicx}
\usepackage{dsfont}
\usepackage[makeroom]{cancel}
\usepackage{url}
\usepackage{hyperref}
\usepackage{slashed}
\usepackage{braket}

\newcommand{\secn}[1]{Section~\ref{#1}}

\newcommand{\beq}{\begin{eqnarray}}
\newcommand{\eeq}{\end{eqnarray}}

\newcommand{\vare}{\varepsilon}

\newcommand{\Gam}{\Gamma}

\newcommand{\lam}{\lambda}
\newcommand{\zbar}{\bar{z}}

\newcommand{\e}{\epsilon}

\def\eq#1{Eq.~(\ref{#1})}

\title{
Non-abelian soft radiation data for a celestial theory
}

\author[a]{Lorenzo Magnea,}
\author[b]{Enrico Zunino}
\affiliation[a]{Dipartimento di Fisica, Universit\`a di Torino, and INFN, Sezione di Torino,\\
                     Via Pietro Giuria 1, I-10125 Torino, Italy}
\affiliation[b]{Higgs Centre for Theoretical Physics, School of Physics and Astronomy,\\
                    The University of Edinburgh, Edinburgh EH9 3FD, Scotland, UK}

\emailAdd{lorenzo.magnea@unito.it}
\emailAdd{enrico.zunino@ed.ac.uk}

\abstract{
Celestial holography posits that the long-distance behavior of gauge and gravity theories
is dictated by two-dimensional conformal field theories defined on the celestial sphere. 
For non-abelian gauge theories, this proposal is verified, to all perturbative orders, by 
dipole color correlations in the infrared factor of non-abelian scattering amplitudes, which 
are given by a correlator of matrix-valued vertex operators in a free-boson theory on the sphere. 
Decades of high-order gauge-theory calculations have provided a number of further results that 
can be used to test and constrain a possible celestial theory: they include explicit expressions
for soft emission currents up to three particles, and up to three loops for single soft emission.
In this paper, we analyze this trove of data, appropriately translated in the celestial language, 
and we use them to extract information on the celestial theory. In particular, we show that all 
logarithms arising in the loop expansion of the single soft current can be reabsorbed in the scale 
choices for the $d$-dimensional coupling, casting some doubt on the need for a logarithmic celestial 
theory. We then note that the celestial OPEs suggested by the structure of multiple emission 
currents in collinear limits are never ambiguous, but involve coefficients depending on gluon 
energy fractions, which break holomorphic factorization, as well as associativity when double 
limits are taken. Strongly-ordered soft limits recover associativity, but suffer from ambiguities 
already discussed in earlier literature.}

\begin{document}
\maketitle


\section{Introduction and motivation}
\label{Intro}

The infrared behavior of gauge-theory scattering amplitudes has been a focus of research
for many decades, beginning with studies that actually precede the formal development of 
quantum field theory~\cite{Bloch:1937pw}. The reason for this long-lived interest is that
the field lies at the intersection of intensely practical concerns (how to compute physically 
relevant cross sections with high precision in the presence of infrared sensitivity) with
deep theoretical questions (the definition of the $S$-matrix, the factorization of physics 
at different length scales, confinement, and the possible holographic behavior of certain
quantum field theories). As a consequence, a wealth of approaches to the problem, and a 
wealth of results, have accrued over the space of decades (for a review, see for 
example~\cite{Agarwal:2021ais}).

A new and radically different perspective on the infrared problem for gauge theories and 
gravity was introduced for the first time in~\cite{Strominger:2013lka}, and rapidly developed
(as reviewed in~\cite{Strominger:2017zoo}). The central idea of this approach is holographic:
even in flat Minkowsky space, infrared information is expected to be encoded in the behavior
of fields on the {\it celestial sphere}, which we can loosely define as the locus of the 
endpoints of all light-like lines emerging from the origin of Minkowsky space. One can then 
argue that infrared properties of scattering amplitudes can emerge from an effective 
theory on the sphere. Since the Lorentz group on the celestial sphere is naturally represented
as $SL(2, {\mathbb C})$, which is the global conformal group in 2 dimensions, one is led
to the exciting conjecture that gauge-theory and gravity amplitudes (or at least their 
infrared limits) could be computed using two-dimensional conformal field theory data, 
which might lead to vast simplifications in perturbative calculations, and possibly to
the extraction of all-order, or even non-perturbative information\footnote{While in general
it is expected that the celestial approach may access non-perturbative information (see, for 
example, Ref.~\cite{Strominger:2017zoo}), we emphasize that our viewpoint in this paper is 
strictly perturbative.}.

The strongest form of this conjecture emerged in Ref.~\cite{Arkani-Hamed:2020gyp}, following
up on earlier work in~\cite{Pasterski:2016qvg,Pasterski:2017kqt}: the suggestion there is to shift 
the focus from conventional scattering amplitudes, expressed in terms of momentum eigenstates, 
to {\it celestial amplitudes}, defined in terms of boost eigenstates, and thus naturally 
forming representations of the (global) conformal group on the sphere. Celestial amplitudes 
are Mellin transforms of conventional amplitudes, trading energy dependence for dependence
on conformal weight. These ideas have been fleshed out in many studies (for reviews, and a 
comprehensive bibliography, see for example~\cite{Pasterski:2021raf,Pasterski:2023ikd}), 
predominantly in the context of gravity theories, but with many results also for QED and 
Yang-Mills theories. Most of these studies concern the structure of tree-level amplitudes (for 
example~\cite{Fan:2019emx, Pasterski:2017ylz, Schreiber:2017jsr}), in some cases pushing the analysis to 
the one-loop level \cite{Gonzalez:2020tpi}, and many explicit results concern {\it soft theorems}, 
analyzing tree-level amplitudes in terms of power expansions in particle energies. These results 
can thus be seen as (semi)-classical, and it is to some extent natural that they should admit a 
holographic interpretation, since they concern long-distance properties of scattering amplitudes. 
It is of great interest therefore to explore the possibility that deep quantum corrections (arising 
at high perturbative orders in the conventional approach) might also be described by a celestial theory.

A step in this direction was taken in Ref.~\cite{Magnea:2021fvy}, exploiting the infrared
factorization properties of non-abelian scattering amplitudes\footnote{Related results in the 
language of {\it celestial amplitudes} were obtained in Ref.~\cite{Gonzalez:2021dxw}.}. In that 
context, one starts by noting that leading-power infrared effects (and thus all infrared divergences) 
are encoded, to all orders in perturbation theory, in a soft color operator acting on a hard, infrared 
finite matching amplitude~\cite{Agarwal:2021ais}. The soft color operator is a natural target for a 
celestial interpretation: it is a long-distance factor from the gauge-theory viewpoint, and yet, 
in the non-abelian case, it has a highly non-trivial perturbative expansion, where deep quantum 
effects play an important role. Ref.~\cite{Magnea:2021fvy} showed that all color-dipole contributions 
to the soft infrared factor, to all orders, can indeed be computed as a correlator of matrix-valued 
vertex operators in a theory of free bosons on the sphere, providing a concrete realization of the 
conjectured celestial CFT. Importantly, color-dipole correlations are not the whole story. At high 
orders in perturbation theory, higher color multipoles appear, starting with quadrupoles at three 
loops~\cite{Almelid:2015jia,Almelid:2017qju}. Furthermore, higher-rank Casimir invariants of the 
gauge algebra come into play, starting at four loops~\cite{Gardi:2009qi, Becher:2019avh}. It would 
be remarkable if these intricate and deeply quantum effects would turn out to be encoded in a 
celestial CFT: in turn, such a celestial theory might become a powerful computational tool.

Results such as those of Refs.~\cite{Almelid:2015jia,Almelid:2017qju} are examples of a
more general fact. Leading-power infrared effects in non-abelian gauge theories are known
to impressive accuracy, thanks to decades of work, largely motivated by phenomenological 
efforts. This applies to both virtual corrections to fixed-angle multi-leg scattering 
amplitudes, and to the currents describing the real radiation of multiple soft particles 
at leading power. All these high-order corrections (reviewed in~\cite{Agarwal:2021ais}) 
provide perturbative data and powerful constraints for a putative celestial theory, but 
have so far received limited attention.

The purpose of our paper is to collect, organize and present some of these high-order 
results in the language of celestial holography, and highlight the evidence for the
existence of an effective celestial theory, as well as the most significant challenges to 
it, pointing to effects that would be very unconventional for a two-dimensional CFT\footnote{It 
is, by now, widely understood that CCFTs must have peculiar characteristics that distinguish them 
substantially from standard textbook CFTs. See, for example, Ref.~\cite{Iacobacci:2024laa}}. We 
will focus here on single and multiple soft emission currents, leaving the analysis of 
multipole virtual corrections to fixed-angle amplitudes to future work. We will begin in 
\secn{Nonabfact} with a very concise review of the factorization of fixed-angle scattering
amplitudes, in order to set up our notation, and we will briefly describe the free-boson
CFT introduced in Ref.~\cite{Magnea:2021fvy}. In \secn{FactRad}, we will discuss the known
results on the factorization of soft radiation from hard scattering amplitudes, at leading 
power in soft energies, and we will introduce our conventions for translating the results 
between the two languages.

\secn{SingSoft} is devoted to the single soft-gluon emission current, which is known for 
massless non-abelian theories up to two loops in full generality~\cite{Dixon:2019lnw}, and 
up to three loops for amplitudes involving only two hard colored particles~\cite{Herzog:2023sgb}. 
Here we discuss the physical significance of logarithmic terms that emerge starting at one loop, 
which turn out to be entirely associated with the scale of the gauge coupling, and we emphasize 
the appearance of more intricate color and kinematic structures starting at two loops. 
In \secn{DoubleSoft}, we turn our attention to the double soft emission current, which is 
known at one loop in full generality~\cite{Zhu:2020ftr, Czakon:2022dwk}. Here the focus is 
on strong-ordering and collinear limits, and helicity dependence. We note, however, that 
gauge-theory data extends beyond these limits, providing constraints that go beyond the reach 
of the conformal operator product expansion (OPE). Finally, in \secn{TripleSoft}, we consider 
the triple soft emission current, which on the gauge-theory side is known only at tree 
level~\cite{Catani:2019nqv}. Even at leading order, however, it contains very non-trivial 
color and kinematic structures. In particular, with three soft emitted particles, it becomes 
possible to take consecutive collinear limits\footnote{Multi-collinear limits of tree-level 
amplitudes have also been investigated in the context of celestial holography in 
Ref.~\cite{Ebert:2020nqf}.}, thus testing the associativity of the corresponding 
conformal OPE in different helicity configurations.


\section{Infrared factorisation for fixed-angle scattering amplitudes}
\label{Nonabfact}

The factorization of infrared divergences in multi-leg fixed-angle non-abelian scattering
amplitude has a long history, starting with work done in the early 1980's~\cite{Sen:1982bt}.
After many refinements over the years (see, for example,~\cite{Botts:1989kf,Dixon:2008gr,
Feige:2014wja}), our current understanding can be summarized as follows. Let ${\cal A}_n$
be the renormalized scattering amplitude for $n$ massless particles in any representation 
of the gauge group, carrying momenta $p_i$. Employing dimensional regularization for infrared
divergences, with $d = 4 - 2 \e$ and $\e < 0$, we can write
\beq
\label{factamplvirt}
  {\cal A}_n \bigg( \frac{p_i}{\mu}, \alpha_s (\mu), \e \bigg) \, = \, 
  {\cal Z}_n \bigg( \frac{p_i}{\mu}, \alpha_s (\mu), \e \bigg) 
  {\cal H}_n \bigg( \frac{p_i}{\mu}, \alpha_s (\mu), \e \bigg) \, .
\eeq
The scattering amplitude ${\cal A}_n$ carries implicit color indices, and can be understood
as a vector in the product space of the color representations of individual particles. The 
infrared factor ${\cal Z}_n$ is a matrix in that space, containing all soft and collinear 
divergences of the amplitude, and it acts on the finite vector ${\cal H}_n$, playing the 
role of a matching function. The infrared factor obeys a renormalization group equation,
which can be solved in terms of path-ordered exponentials (denoted below by $P$) of a 
matrix-valued anomalous dimension. This yields
\beq
\label{solZ}
  {\cal Z}_n \bigg( \frac{p_i}{\mu}, \alpha_s (\mu), \e \bigg) \, = \, 
  P \exp \left[ \frac{1}{2} \int_0^{\mu^2} \frac{d \lambda^2}{\lambda^2} \,
  \Gamma_n \bigg( \frac{p_i}{\lambda}, \alpha_s (\lambda, \e) \bigg) \right] \, .
\eeq
Note that the {\it soft anomalous dimension} $\Gamma_n$ is a finite function of its 
arguments, and all infrared poles of  the amplitude are generated by performing the integration 
over the scale $\lambda$ of the $d$-dimensional running coupling $\alpha_s(\lambda, \e)$, which
in turn obeys
\beq
\label{beta}
  \lambda \frac{\partial \alpha_s}{\partial \lambda} \, \equiv \, \beta(\alpha_s, \e) \, = \, 
  - 2 \e \alpha_s - \frac{\alpha_s^2}{2 \pi} \, \sum_{k = 0}^\infty \bigg( \frac{\alpha_s}{\pi}
  \bigg)^{\! k} b_k \, ,
\eeq
with $b_0 = 11 C_A/3$ for pure Yang-Mills theory in our normalization.

In order to give explicit expressions for $\Gamma_n$, one must choose a notation for 
color operators. One way to do that is to pick a basis of tensors in color space, and 
a common choice in this case is to pick the (overcomplete) basis given by traces of
color generators in the fundamental representation. When dealing with all-order 
properties for an arbitrary number of particles, however, we argue here that the
most powerful notation uses the {\it color insertion operators} ${\bf T}_i$ ($i = 1, 
\ldots, n)$ introduced and popularized in \cite{Bassetto:1984ik,Catani:1996vz}. The 
operators ${\bf T}_i$ act on the tensor product space ${\cal V}_n \equiv \otimes_{p = 1}^n
V_p$, where $V_p$ is the color representation of particle $p$: they act non-trivially 
only on the $i$-th factor of the product, where they are represented by the appropriate
generator of the gauge algebra. As such, they obey
\beq
\label{propT}
  \Big[ {\bf T}_i^a, {\bf T}_j^b \Big] \, = \, {\rm i} f^{ab}_{\,\,\,\,\,\, c} \, 
  {\bf T}_i^c \, \delta_{ij} \, ,
  \qquad 
  {\bf T}_i \cdot {\bf T}_i \, \equiv \, {\bf T}_i^a {\bf T}_i^b \, \delta_{ab} 
  \, = \, C_i \, ,
  \qquad
  \sum_{i = 1}^n {\bf T}_i \, = \, 0 \, , 
\eeq 
where $C_i$ is the quadratic Casimir eigenvalue of the algebra ($C_A$ for the 
adjoint representation, for example), and the last identity, embodying gauge invariance, 
is understood to be valid when color operators are acting on gauge-invariant
quantities. This notation has the advantages of being independent of the choice of 
basis, as well as independent of the representation content of the amplitude (the 
last equality in \eq{propT} simply enforces the fact that the amplitude is a color 
singlet, or, in other words, the decomposition of the space ${\cal V}_n$ in irreducible
representations must contain the singlet representation). Finally, this notation treats
uniformly the number of hard particles. From a physical viewpoint, the operators ${\bf T}_i$
represent (matrix-valued) charges, and in many instances one can upgrade QED expression
to the non-abelian case by simply replacing electric charges with color operators. 

To illustrate the power of the color operator notation, we briefly summarize what is 
known about the soft anomalous dimension matrix $\Gamma_n$. In the massless case it can 
be written as
\beq
\label{AllGamma}
  \Gamma_n \bigg( \frac{p_i}{\mu}, \alpha_s(\mu) \bigg) \, = \, 
  \Gamma_n^{\rm \, dip} \bigg( \frac{s_{ij}}{\mu^2}, \alpha_s(\mu) \bigg) +
  \Delta_n \Big( \rho_{ijkl}, \alpha_s (\mu) \Big) \, ,
\eeq
up to corrections proportional to higher-order Casimir eigenvalues, starting at four loops. 
In \eq{AllGamma}, the first term encodes all color-dipole contributions, and it is given 
by~\cite{Gardi:2009qi,Becher:2009cu,Becher:2009qa,Gardi:2009zv}
\beq
\Gamma_n^{\rm \, dipole} \bigg( \frac{s_{ij}}{\mu^2}, \alpha_s(\mu) \bigg) \, = \,
  \frac{1}{2} \, \widehat{\gamma}_K \big( \alpha_s(\mu) \big) \sum_{k = 1}^n 
  \sum_{l = k + 1}^n \log  \frac{- s_{kl} + {\rm i} \eta}{\mu^2} \,\,\, 
  {\bf T}_k \cdot {\bf T}_l 
  \, - \, \sum_{k = 1}^n \gamma_k \big( \alpha_s (\mu) \big) \, .
\eeq
Here $\widehat{\gamma}_K$ gives the Casimir-scaling part of the cusp anomalous dimension,
which, up to three loops, is related to the full cusp in representation $r$ by
\beq
\label{Cascal}
  \gamma_{K, r} (\alpha_s) \, = \, C^{(2)}_r \, \widehat{\gamma}_K (\alpha_s) \, ,
\eeq
while $\gamma_i$ is the collinear anomalous dimension for particle $i$. These anomalous 
dimensions are known to four loops (see, for example,~\cite{vonManteuffel:2020vjv}).
For completeness, although it will not be further discussed here, we note that the 
second term in \eq{AllGamma} is constrained to depend only on scale-invariant cross 
ratios of momenta (a very promising feature for a possible celestial CFT): specifically
\beq
\label{Cicr}
  \rho_{ijkl} \, = \, \frac{p_i \cdot p_j \, p_k \cdot p_l}{p_i \cdot p_k \, p_j \cdot p_l} 
  \, = \, \frac{s_{ij} s_{kl}}{s_{ik} s_{jl}} \, .
\eeq
The color operator $\Delta_n$ is known explicitly at three loops, where it is built
out of color quadrupole correlations. It is given by~\cite{Almelid:2015jia}
\beq
\label{Delta3}
        &&
        \Delta^{(3)}_n(\rho_{ijkl}) \, = \, \frac{1}{4} f_{a b e} f^e_{\phantom{e} c d}
        \Biggl\{- \, C \, \sum_{i=1}^n \sum_{\substack{1 \le j < k \le n \\ j,k \ne i}}
        \{\mathbf{T}_i^a, \mathbf{T}_i^d\} \mathbf{T}_j^b \mathbf{T}_k^c \nonumber \\
        && \hspace{1cm} + \, \sum_{1 \le i < j < k < l \le n}
        \biggl[ \mathbf{T}_i^a \mathbf{T}_j^b \mathbf{T}_k^c \mathbf{T}_l^d \,
        \mathcal{F}(\rho_{ikjl}, \rho_{iljk}) \, + \, 
        \mathbf{T}_i^a \mathbf{T}_k^b \mathbf{T}_j^c \mathbf{T}_l^d \,
        \mathcal{F}(\rho_{ijkl}, \rho_{ilkj}) \nonumber \\
        && \hspace{3.5cm} + \, \mathbf{T}_i^a \mathbf{T}_l^b \mathbf{T}_j^c \mathbf{T}_k^d \, 
        \mathcal{F}(\rho_{ijlk}, \rho_{iklj}) \biggr] \Biggr\} \,.
\eeq
The constant $C = \zeta_5 + 2 \zeta_2 \zeta_3$ appearing in the first line of \eq{Delta3} 
is crucial to preserve the factorization properties of amplitudes in collinear limits 
\cite{Almelid:2015jia}. The kinematic dependence, encoded by the single function ${\cal F}$,
is remarkably simple, and is given by a combination of {\it single-valued harmonic 
polylogarithms} (SVHPLs)~\cite{Brown:2004ugm} of uniform weight 5; the sum over 
color-quadrupoles, together with the kinematic dependence, is constructed to enforce 
Bose symmetry, as required.

Regarding a possible celestial interpretation of \eq{factamplvirt}, our viewpoint is that
the the infrared factor ${\cal Z}_n$ is a natural candidate for a holographic representation.
Indeed, the color-correlated part of ${\cal Z}_n$ arises from the {\it soft operator}
\beq
\label{softfun}
  {\cal S}_n \left(\beta_i \cdot \beta_j, \alpha_s (\mu), \e \right) \, = \, 
  \bra{0} T \left[ \prod_{k = 1}^n \, \Phi_{\beta_k} (\infty,0 ) \right] \ket{0} \, ,
\eeq
where $\beta_i$ are the four-velocities of the hard particles ($p_i = \mu \beta_i$), and
$\Phi_{\beta_i}$ are semi-infinite straight Wilson lines aligned with the classical particle 
trajectories\footnote{Note that, in keeping with the perturbative approach pursued in this paper, 
the vacuum expectation value in \eq{softfun} is to be evaluated order by order in perturbation 
theory, by expanding the Wilson lines in powers of the coupling. We are thus not accessing 
topologically non-trivial sectors of the theory: rather, we treat \eq{softfun} as a generator 
of perturbative infrared singularities.}. The soft operator therefore is naturally expressed in 
terms of $n$ marked points on the celestial sphere, where color charges represented by ${\bf T}_i$ 
are located. This viewpoint was pursued in Ref.~\cite{Magnea:2021fvy}. There, the 
color-correlated part of the soft factor ${\cal Z}_n$, restricted to color dipoles, was 
written in celestial coordinates as 
\beq
\label{softcorrZn}
  {\cal Z}_n^{\rm corr.} \Big( z_{ij}, \alpha_s (\mu^2), \e \Big) \, = \,
  \exp \left[ - \, K \left( \alpha_s (\mu), \e \right) \, \sum_{k = 1}^n \sum_{l = k+1}^n
  \ln \Big( \left| z_{kl} \right|^2 \Big) \, {\bf T}_k \cdot {\bf T}_l \right] \, ,
\eeq
where, in celestial coordinates, $|z_{ij}|^2 \equiv |z_i - z_j|^2 = \beta_i \cdot 
\beta_j$. The prefactor $K(\alpha_s, \e)$, carrying all coupling, scale, and regulator 
dependence, is the scale average of the universal cusp anomalous dimension, which can
also be understood to play the role of the strong coupling in the infrared (see, for example, 
Refs.~\cite{Catani:1990rr, Grozin:2015kna, Catani:2019rvy}). It is
given by
\beq
\label{Suda}
  K \left( \alpha_s (\mu), \e \right) \, = \, - \, \frac{1}{2} \int_0^\mu 
  \frac{d \lambda}{\lambda} \,\, \widehat{\gamma}_K \big( \alpha_s(\lambda, \e) \big) \, .
\eeq
The main result of Ref.~\cite{Magnea:2021fvy} is that \eq{softcorrZn} is reproduced
by computing a correlator of vertex operators in a celestial CFT of free-bosons,
taking values in the Lie algebra of the gauge group. The action for this theory is simply
\beq
\label{freebos}
  S(\phi) \, = \, \frac{1}{2 \pi} \int d^2 z \, \partial_z \phi^a (z, \bar{z}) \,  
  \partial_{\bar{z}} \phi_a (z, \bar{z}) \, ,
\eeq
while the relevant vertex operator can be defined by
\beq
\label{vertop}
  V(z, \bar{z}) \, = \, : {\rm e}^{{\rm i} \kappa {\bf T} \cdot \phi (z, \bar{z})} : \, .
\eeq
In fact, one finds that, up to an overall normalization, 
\beq
\label{corrcorr}
  {\cal C}_n \big( \{ z_i \}, \kappa \big) \, \equiv \, \bigg< \prod_{k=1}^n
  V (z_k, \bar{z}_k) \bigg>
  \, = \, {\cal Z}_n^{\rm corr.} \Big( z_{ij}, \alpha_s (\mu^2), \e \Big) \, ,
\eeq
provided one identifies the square of the field normalization, $\kappa^2$, with the integral
in \eq{Cascal}, which implies that, to leading order, $\kappa \sim g_s$ plays the role of 
the gauge coupling. Note that \eq{corrcorr} holds up to corrections proportional to the 
structure constants $f_{abc}$ and arising at higher orders in $\kappa$, which can emerge
from the underlying theory, when interaction terms are added to \eq{freebos}. As noted 
above, such corrections do arise in the bulk gauge theory starting at three loops, and 
we leave their discussion to future work.


\section{Soft factorization for real radiation}
\label{FactRad}

The infrared factorization of virtual corrections to fixed-angle scattering amplitudes, 
embodied in \eq{factamplvirt}, must have counterparts for the real radiation of soft and 
collinear particles: indeed, the KLN theorem~\cite{Kinoshita:1962ur,Lee:1964is} dictates 
that the integration of real radiation must provide infrared poles canceling the ones 
present in virtual corrections, and it is hard to envisage such a cancellation without 
a factorization theorem for real radiation. Here, as was the case in \eq{softcorrZn}, 
we focus on {\it soft} radiation, which is the source of all color correlations, and
we will not consider hard collinear radiation, which can always be factorized in 
color-singlet form.

In order to describe the expected factorization, following Refs.~\cite{Magnea:2018ebr,
Magnea:2024jqg}, we introduce the {\it eikonal form factors}
\beq
\label{eikff}
  {\bf S}^{\lambda_1 \ldots \lambda_m} \big(q_1, \ldots, q_m; \{\beta_i\} \big) 
  \, \equiv \, \bra{ \{ q_j, \lambda_j \} } T \left[ \prod_{k = 1}^n \, \Phi_{\beta_k} 
  (\infty,0 ) \right] \ket{0} \, ,
\eeq
which describe the radiation of $m$ soft gluons with momenta $q_j$ and polarizations 
$\lambda_j$ ($j = 1, \ldots, m$) from the system of Wilson lines introduced in \eq{softfun}. 
The eikonal form factors define multiple {\it soft gluon currents}, according to
\beq
\label{multsgc}
  {\bf S}^{\lambda_1 \ldots \lambda_m} \big(q_1, \ldots, q_m; \{\beta_i\} \big) 
  \, = \, \varepsilon^{\lambda_1}_{\mu_1} (q_1) \ldots \varepsilon^{\lambda_m}_{\mu_m} 
  (q_m) \,\, {\bf J}^{\mu_1 \ldots \mu_m} \big(q_1, \ldots, q_m; \{ \beta_i \} \big) \, .
\eeq
Note that this definition in terms of matrix elements of an operator is valid to all
orders in perturbation theory and not only at tree level. Thus eikonal form factors and 
soft gluon currents can be expanded in powers of the strong coupling, as
\beq
\label{pertexpS}
  {\bf S}^{\lam_1 \ldots \lambda_m} \big( q_1, \ldots, q_m; \{\beta_i\} \big)
  \, \equiv \, g_s \sum_{\ell = 0}^{\infty} \left(\frac{\alpha_s}{4 \pi}\right)^\ell 
  {\bf S}^{\lam_1 \ldots \lambda_m, \, (\ell)} \big(q_1, \ldots, q_m; \{\beta_i\} \big) \, .
\eeq
Scattering amplitudes involving $n$ colored hard particles with momenta $p_i = \mu \beta_i$ (and 
any number of colorless particles) and $m$ real soft gluons with momenta $q_j$ and polarizations 
$\lambda_j$ are expected to factorize, at {\it leading power} in all the soft energies, as
\beq
  {\cal A}_{\, n + m}^{\lam_1 \ldots \lambda_m} \bigg( q_1, \ldots, q_m; \frac{p_i}{\mu} 
  \bigg) \, = \, 
  {\bf S}^{\lam_1 \ldots \lambda_m} \big( q_1, \ldots, q_m; \{\beta_i\} \big) \,  
  {\cal A}_n \bigg( \frac{p_i}{\mu} \bigg) \, ,
\label{factamplmultreal}
\eeq
where the dependence on the strong coupling and on $\e$ is understood, as are the color indices. 
The soft operators carry the color indices of the soft gluons, as will become clear below. 
\eq{factamplmultreal} has been shown to hold by explicit calculations for single gluon emission 
up to two loops \cite{Catani:1998bh,Dixon:2019lnw} (and up to three loops in the case $n=2$ in 
Ref.~\cite{Herzog:2023sgb}), for double gluon emission up to one loop~\cite{Zhu:2020ftr,Czakon:2022dwk}, 
while at tree level a formal proof was given for any $m$ in \cite{Catani:1999ss}, and explicit 
calculations have been carried out up to $m=3$~\cite{Catani:2019nqv} for any $n$. It is generally 
assumed that the leading-power factorization in \eq{factamplmultreal} holds for any $m$ and $n$, 
and to all orders in perturbation theory.

To illustrate these statements, we now consider explicitly the case $m=1$ at tree level,
and we use this simple example to introduce our notations to represent the currents on 
the celestial sphere. For single-gluon radiation, \eq{factamplmultreal} reads
\beq
\label{factamplsingreal}
  {\cal A}_{\, n + 1}^\lam \bigg( q; \frac{p_i}{\mu} \bigg) \, = \, 
  {\bf S}^\lam \big( q; \{\beta_i\} \big) \,  
  {\cal A}_n \bigg( \frac{p_i}{\mu} \bigg) 
  \, = \, 
  \varepsilon^{\lam} (q) \cdot {\bf J} \big( q, \{\beta_i\} \big) \, 
  {\cal A}_n \bigg( \frac{p_i}{\mu} \bigg) \, .
\eeq
Expanding the single-gluon eikonal form factor according to \eq{pertexpS}, a
straightforward calculation yields the well-known expression
\beq
\label{softglu0}
  {\bf J}^{\mu,(0)} (q) \, = \,  \sum_{i = 1}^n  \, 
  \frac{\beta_i^\mu}{\beta_i \cdot q} \, \, {\bf T}_i \, ,
\eeq
where for simplicity we dropped the dependence on $\beta_i$ in the arguments of 
the current. Importantly, the tree-level current is {\it gauge invariant}: indeed it satisfies
\beq
  q \cdot  {\bf J}^{\mu,(0)} (q) \, =  \, \sum_{i = 1}^n \, {\bf T}_i \, = \, 0 \, ,
\label{gaugeinv}
\eeq
where the last equality holds when the current acts as a color operator on the
gauge-invariant Born amplitude, as in \eq{factamplsingreal}.

To translate the representation of the current in terms of four-momenta, \eq{softglu0}, to 
celestial coordinates, we use the standard parametrization of momenta
\beq
\label{celparam}
    p_i^{\mu} \, = \, \frac{\omega_i}{\sqrt{2}} \, \Big\{1 + z_i \zbar_i, z_i + \zbar_i, - 
    {\rm i} (z_i - \zbar_i), 1 - z_i \zbar_i \Big\} \, .
\eeq
Similarly, we will parametrize the soft momentum $q$ as
\beq
\label{celparamq}
    q^{\mu} \, = \, \frac{\omega_q}{\sqrt{2}} \, \Big\{1 + z \zbar, z + \zbar, - 
    {\rm i} (z - \zbar), 1 - z \zbar \Big\} \, .
\eeq
We work in a regime where all $\omega_i$ are parametrically of the same order, while
$\omega_q \ll \omega_i$. All the results that we are discussing are valid up to corrections 
suppressed by powers of $\omega_q/\omega_i$. In this regime, at leading power, soft radiation
becomes insensitive to the precise values of the energies $\omega_i$, which can be scaled 
out, and soft currents can be expressed in a scale-invariant way in terms of the four-velocities
associated with the hard momenta $p_i$, which we define by
\beq
\label{defbet}
  p_i^{\mu} \, = \, \omega_i \beta_i^{\mu} \, .
\eeq
The final necessary ingredient is the expression for the gluon polarisation vectors in 
celestial coordinates. As is usual, we write
\beq
  \vare^{\mu}_+ (q) & = & \frac{1}{\omega_q} \, \partial_zq^\mu \, = \,
  \frac{1}{\sqrt{2}} \, \big\{ \zbar, 1, - {\rm i}, - \zbar \big\} \, , 
  \nonumber \\
  \vare^{\mu}_- (q) & = & \frac{1}{\omega_q} \, \partial_{\zbar} q^\mu \, = \, 
  \frac{1}{\sqrt{2}} \, \big\{ z, 1, {\rm i}, - z \big\} \, ,
\eeq
satisfying
\beq
\label{polarnorm}
  \left(\vare_\pm (q) \right)^2 \, = \, 0 \, , \qquad
  \vare_+ (q) \cdot \vare_- (q) \, = \, -1 \, , \qquad
  q \cdot \vare_\pm (q) \, = \, 0 \, .
\eeq
With these definitions, it is easy to obtain expressions for the relevant scalar products 
in celestial coordinates. One finds
\beq
\label{celscalprod}
  p_i \cdot p_j \, = \, \omega_i \omega_j|z_{ij}|^2 \, , \qquad \!
  \beta_i \cdot \beta_j \, = \, |z_{ij}|^2 \, , \qquad \!
  \vare_+(q) \cdot p_i \, = \, \omega_i \zbar_{qi} \, , \qquad \!
  \vare_-(q) \cdot p_i \, = \, \omega_i z_{qi} \, , \hspace{5mm} 
\eeq
where we introduced the notation $z_{ab} = z_a - z_b$, and similarly $\zbar_{ab} = 
\zbar_a - \zbar_b$. At tree-level, applying our dictionary to \eq{softglu0} results in 
\beq
\label{treesoftcurr}
  {\bf S}^{+, (0)} (q) \, = \, \sum_{i = 1}^n \, \frac{\vare_+ \cdot \beta_{i}}{q \cdot \beta_i} 
  \,\, {\bf T}_i \, = \, \frac{1}{\omega_q} \, \sum_{i = 1}^n \, \frac{1}{z_{qi}} \,\,  {\bf T}_i 
  \, ,\\
  {\bf S}^{-, (0)} (q) \, = \, \sum_{i = 1}^n \, \frac{\vare_- \cdot \beta_{i}}{q \cdot \beta_i}
  \,\, {\bf T}_i \, = \, \frac{1}{\omega_q} \, \sum_{i = 1}^n \, \frac{1}{\zbar_{qi}} 
  \,\, {\bf T}_i \, .
\eeq
On the celestial sphere, these tree-level currents emerge from the Ward identities for the
conserved Noether currents associated with the symmetry of the action in \eq{freebos} 
under translations in field space~\cite{Strominger:2013lka,Pate:2019mfs,Pasterski:2021raf,
Pasterski:2021rjz,Magnea:2021fvy}. The currents are simply
\beq
\label{NoethCurr}
  j^a (z) \, = \, \partial_z \phi^a (z, \bar{z}) \, , \qquad \quad 
  \bar{j}^a (\bar{z}) \, = \, \partial_{\bar{z}} \phi^a (z, \bar{z}) \, ,
\eeq
and they are, respectively, holomorphic and anti-holomorphic weight-1 primary fields in the 
theory defined by \eq{freebos}. Their conservation ({\it i.e.} their (anti-)holomorphicity)
entail the Ward identities
\beq
\label{WId}
  \Big\langle j^a (z) \, \prod_{i=1}^n V (z_i, \bar{z}_i) \Big\rangle \, = \, 
  - \, \frac{{\rm i}}{2} \, \sum_{j = 1}^n \frac{{\bf T}^a_j}{z - z_j} \, 
  {\cal C}_n \big( \{ z_j \}, \kappa \big) \, ,
\eeq
and similarly for the anti-holomorphic current. This lets us interpret \eq{WId}, and its 
complex conjugate, as the celestial versions of \eq{factamplsingreal}. To be specific, note 
that \eq{factamplsingreal} applies also when we replace the full amplitude ${\cal A}_n$ on 
the right-hand side with its infrared factor ${\cal Z}_n$, and we correspondingly replace 
the $n$ hard particles on the left-hand side with their Wilson lines. \eq{WId} then matches 
\eq{factamplsingreal}, provided we interpret $j^a(z)$ as a dimensionless operator creating a 
soft gluon with positive polarization and color index $a$, while the anti-holomorphic current 
similarly creates a soft gluon with negative polarization. Since the conformal current must be 
dimensionless, it generates the soft factors in \eq{treesoftcurr} multiplied times the soft 
energy $\omega_q$. The remaining singularities as $z \to  z_i$ are then the collinear singularities 
emerging as the soft momentum $q$ becomes parallel to the Wilson line along $\beta_i$.

There are a few things to remark about this correspondence. First of all, we note that the 
collinear singularities in \eq{WId} can be seen to arise from the OPE of the celestial current 
with each one of the vertex operators. The OPE would however admit holomorphic corrections,
{\it i.e.} contributions that are not singular as $z \to z_i$. These contributions vanish
in \eq{WId}, when the OPE is taken inside the vacuum expectation value, so that \eq {WId}
is exact. Correspondingly, the single soft-gluon current has no terms without collinear
singularities. Another interesting fact is that the correlator ${\cal C}_n$ reproduces 
the soft factor ${\cal Z}_n$ to {\it all orders} in perturbation theory (for dipole 
correlations), while the Ward identity in \eq{WId} yields the soft gluon current only 
at {\it tree level}. This would be the exact result in QED\footnote{Note, however, that
even in QED the single soft-photon current receives corrections, starting at three loops,
in the presence of massless charged particles~\cite{Ma:2023gir}.}, but in the non-abelian 
theory there are loop corrections which, at least in pure Yang-Mills theory, are entirely 
due to gluon self-interactions, and are proportional to the structure constants $f_{abc}$.
These corrections will be discussed in the next section. They cannot be controlled by
the free-boson theory defined by \eq{freebos}, which must turn into an interacting
two-dimensional theory with interaction terms proportional to the structure constants.
Loop corrections to soft-gluon currents thus must provide important pointers to build 
the interacting celestial theory.


\section{The single-emission current}
\label{SingSoft}

In this section, we explore the known loop corrections to the tree-level soft current representing 
the emission of a single gluon in a massless gauge theory. We translate these results in celestial 
coordinates and analyze the structures that arise up to two loops. As we will see, these gauge-theory
data contain useful information for the construction of the putative CCFT that should mirror the 
gauge theory on the celestial sphere. 


\subsection{The single-emission current at one loop}
\label{Oloone}

Let us begin with the first quantum correction to the tree-level soft current in \eq{softglu0}. 
This was computed for the first time in the trace basis in Refs.~\cite{Bern:1998sc,Bern:1999ry}, 
and in the present basis-independent notation in Ref.~\cite{Catani:2000pi}. With our conventions, 
it reads
\beq
\label{Cata1}
  {\bf S}_a^{\lam, (1)} (q) \, = \, {\rm i} C_1 (\e) f_{abc}  \sum_{i \ne j}^n \, {\bf T}^b_i
  {\bf T}_j^c \, \left( \frac{\vare^{\lam} \cdot \beta_i}{\beta_i \cdot q} - \frac{\vare^{\lam} 
  \cdot \beta_j}{\beta_j \cdot q} \right) \left(\frac{\mu^2 s_{ij}}{s_{qi} s_{qj}} \right)^\e \, ,
\eeq
where the prefactor, in the $\overline{{\rm MS}}$ regularization scheme, is given by
\beq
\label{C1}
  C_1(\e) \, = \, - \frac{e^{\e \gamma_E}}{\e^2} \frac{\Gam^3(1 - \e) 
  \Gam^2(1 + \e)}{\Gam(1-2\e)} \, = \, - \frac{1}{\e^2} - \frac{\zeta(2)}{2} + 
  \frac{7}{2}\zeta(3) \e + {\cal O}(\e^2) \, . 
\eeq
This result deserves several comments.
\begin{itemize}
  \item It is important to underline that the structure of \eq{Cata1} is completely
  determined by power counting, gauge invariance, Bose symmetry, and rescaling invariance
  for the four-velocities $\beta_i$: the only factor requiring an actual calculation is
  the numerical prefactor $C_1(\e)$, whose Laurent expansion in powers of $\e$ must start
  with a double pole because of the overlapping soft and collinear singularities arising
  in the one-loop calculation.
  \item As announced in \secn{FactRad}, \eq{Cata1} is purely non-abelian, and it 
  vanishes in QED. The color structure suggests that this correction cannot arise in 
  the CCFT from the free-boson action in \eq{freebos}: rather, it will emerge from an
  interacting CCFT.
  \item We note that \eq{Cata1} is the {\it bare} single soft current, directly 
  extracted from Feynman diagram calculations. The renormalization properties of soft currents 
  are readily deduced from their definition in \eq{factamplmultreal}. For on-shell massless gauge 
  amplitudes, only the coupling needs to be renormalized: as a consequence, the $m$-gluon soft 
  current must renormalize multiplicatively, with $m$ powers of the renormalization constant 
  for the coupling, $Z_g$. This means that the renormalized single soft current at one loop differs 
  from the bare one in \eq{Cata1} just by the addition of a term proportional to tree-level
  current, \eq{softglu0}, and proportional to the first $\beta$-function coefficient $b_0$.
  \item Expanding \eq{Cata1} in powers of $\e$, as we do explicitly below, one finds
  (unsurprisingly) that the one-loop correction to the current contains logarithms of
  the eikonal factor $\mu^2 s_{ij}/(s_{ik} s_{jk})$. This has been taken to suggest that the
  CCFT should be a {\it logarithmic} CFT \cite{Bhardwaj:2024wld, Bissi:2024brf, Fiorucci:2023lpb, 
  Kogan:1997fd, Flohr:2001zs, Creutzig:2013hma}. However, as discussed below, the logarithms 
  generated by the eikonal factor have a straightforward interpretation in the gauge theory as 
  {\it ultraviolet} logarithms, naturally associated with the scale of the running coupling. One 
  may then wonder if the logarithms should be attributed to the CCFT, or taken as a physical 
  input from the bulk theory.
\end{itemize}

\noindent
The first step is obviously to rewrite \eq{Cata1} in celestial coordinates. Picking
for example the emission of a gluon with positive helicity, we find
\beq
\label{CelCurr1lo}
  \omega_q {\bf S}_a^{+, (1)} \, = \, {\rm i} C_1(\e) \left(\frac{\mu^2}{2 \omega_q^2} \right)^{\! \e}
  f_{abc}  \sum_{i \ne j}^n {\bf T}^b_i {\bf T}_j^c \,\, \frac{z_{ij}}{z_{qi}z_{qj}} \,
  \left| \frac{z_{ij}}{z_{qi}z_{qj}} \right|^{2 \e} \, ,
\eeq
while the result for a gluon of negative helicity is simply obtained by swapping non-barred with
barred variables. In \eq{CelCurr1lo}, we have chosen to multiply the eikonal form factor times
the light-cone energy of the soft gluon, $\omega_q$, so that the {\it r.h.s.} is dimensionless 
in $d=4$. This is the only kind of object that might emerge from a CCFT, and indeed the Ward 
identity in \eq{WId} generates precisely $\omega_q {\bf S}_a^{(0),\pm}$. \eq{CelCurr1lo} is however
still problematic from a CCFT point of view, for several reasons: first, there is residual 
scale and energy dependence in $d = 4 - 2 \e$, which will survive in the $\e$-expansion due to
the overall $\e^{-2}$ soft-collinear singularity; second, as mentioned above, upon expanding 
in powers of $\e$, one finds logarithmic dependence on the celestial coordinates $z_i$; finally,
the expression in \eq{CelCurr1lo} is neither holomorphic nor antiholomorphic, which again seems
to call for a rather unconventional CCFT.

Beginning with the energy and regulator dependence, we follow the logic in Ref.~\cite{Magnea:2021fvy}:
there, it was noted that the free-boson CCFT predicts only the color-correlated part of the soft 
factor, which is energy-independent, while energy dependence is confined to collinear factors
that lie outside the scope of the CCFT. It is easy to see that the one-loop soft current follows
a similar, though not identical, pattern. For example, the double pole in \eq{CelCurr1lo} is
of soft-collinear origin, and one can easily see that it does not involve color correlations
between hard particles: rather, it is a sum of emissions from individual legs. In order to make 
this explicit, we define the shorthand symbols
\beq
\label{Shand}
  \mathcal{Z}_{ij}^q \, = \, \frac{z_{ij}}{z_{qi}z_{qj}} \, = \, \frac{1}{z_{qi}} - \frac{1}{z_{qj}} 
  \, , \qquad \quad 
  \Omega_q \, = \, \frac{\mu^2}{2 \omega_q^2} \, .
\eeq
Expanding \eq{CelCurr1lo} in powers of $\e$, we now get
\beq
\label{CelCurr1loExp}
  \omega_q {\bf S}_a^{+, (1)} & = & {\rm i} f_{abc} \sum_{i \ne j}^n
  {\bf T}^b_i {\bf T}_j^c \, \mathcal{Z}_{ij}^q \, \bigg[ \left( - \frac{1}{\e^2} - 
  \frac{1}{\e} \log \left| \mathcal{Z}_{ij}^q \right|^2 - 
  \frac{1}{2} \log^2 \left|\mathcal{Z}_{ij}^q \right|^2 - \frac{\zeta_2}{2} \right) + \nonumber \\
  && \hspace{5mm} + \, \log \Omega_q \left( - \frac{1}{\epsilon} - \log \left| \mathcal{Z}_{ij}^q 
  \right|^2 \right) - \log^2 \Omega_q \bigg] + {\cal O} (\e) \, .
\eeq
Here, the coefficients of the powers of $\log\Omega_q$ in the expansion are equivalent, once we 
account for differences in the notation, to the conformally soft operators defined at tree-level 
in Refs.~\cite{Himwich:2021dau,Guevara:2021abz}, and extended to the one-loop case in 
Ref.~\cite{Bhardwaj:2024wld}.  

Next, we notice that, for all terms whose kinematic factor does not depend on the hard particle 
labels $i$ and $j$, and for those terms that depend only on one of the two labels, we can perform 
at least one of the two sums over $i$ and $j$ using color conservation. For example
we find
\beq
\label{ColCons}
  f_{abc} \sum_{i \ne j}^n {\bf T}^b_i {\bf T}_j^c \, \mathcal{Z}_{ij}^q & = & 
  f_{abc} \sum_{i \ne j}^n {\bf T}^b_i {\bf T}_j^c \, \left( \frac{1}{z_{qi}} - \frac{1}{z_{qj}} 
  \right) \nonumber \\ 
  & = & f_{abc} \left[ \, \sum_{i = 1}^n \frac{{\bf T}^b_i}{z_{qi}} \,\, 
  \sum_{\substack{j = 1 \\  j \neq i}}^n {\bf T}_j^c \, - \, 
  \sum_{j = 1}^n \frac{{\bf T}^c_j}{z_{qj}} \,\, 
  \sum_{\substack{i = 1 \\  i \neq j}}^n {\bf T}_i^b \right] \\
  & = & - f_{abc} \sum_{i = 1}^n \frac{1}{z_{qi}} \left( {\bf T}^b_i {\bf T}_i^c - 
  {\bf T}^c_i {\bf T}_i^b \right) \, = \, - {\rm i} f_{abc} f^{bc}_{\phantom{bc} d} \sum_{i = 1}^n 
  \frac{{\bf T}_i^d}{z_{qi}} \, = \, \omega_q C_A \, {\bf S}^{+, (0)}_a \, , \nonumber 
\eeq
where we applied color conservation to the second line, keeping in mind the fact that this
can be done only when color operators act directly on a gauge invariant hard amplitude, so
they have to be ordered accordingly.

We can now organize the current as the sum of a color-uncorrelated term, proportional to 
the tree-level current, and a purely non-abelian, color-correlated contribution, writing 
\beq
\label{CorrUncorr}
  \omega_q{\bf S}_a^{+,(1)} & = & {\rm i} \biggl( - \frac{1}{\epsilon^2} - 
  \frac{1}{\epsilon} \log \Omega_q - \frac{\zeta_2}{2} - \log^2\Omega_q \biggr) \,
  C_A \, {\bf S}_a^{+,(0)}\nonumber \\
  & + &
  {\rm i} \sum_{i \ne j} f_{abc} {\bf T}_i^b {\bf T}_j^c \, \mathcal{Z}_{ij}^q \,
  \biggl[ \biggl(- \frac{1}{\epsilon} - \log \Omega_q \biggr) 
  \log \left| \mathcal{Z}_{ij}^q \right|^2 - 
  \log^2 \left| \mathcal{Z}_{ij}^q \right|^2\biggr] + 
  \mathcal{O} (\epsilon) \, .
\eeq
Note that, to this order in $\e$, the expansion of the prefactor $C_1(\e)$ is entirely associated
with the color-uncorrelated term. Note also that renormalizing the one-loop current would have the
only effect of adding a term proportional to $b_0$ to the parenthesis on the first line.

One aspect of \eq{CorrUncorr} that is problematic from the point of view of a CCFT is the fact
that energy dependent terms are still present in the color-correlated contributions, in the form
of logarithms of the ratio $\Omega_q$, starting at ${\cal O}(\e^0)$. This is in contrast to the 
situation for virtual corrections, as seen for example in \eq{softcorrZn}: there, all dependence
on the coupling, the scale and the regulator was confined to a prefactor that could be matched by 
the field normalization in the vertex operator. We are now going to make an important observation,
which brings the situation for real soft radiation closer to that observed for virtual corrections: 
the basic difference between the two cases is the fact that the tipical energy scale for virtual
corrections is set by the renormalization scale $\mu$, or equivalently by the ultraviolet cutoff,
which is naturally related to the hard scale of the original scattering problem, before the soft
factorization. The natural scale of the soft radiation, instead, is set by the energy (or more
precisely by the transverse momentum) of the radiated gluon. This should be reflected by the
scale choice for the renormalized coupling.

In order to make this observation precise, let's go back to \eq{Cata1} and consider, for 
each color dipole, the scale ratio that is raised to the power $\e$. It is well known (see for 
example Refs.~\cite{Bassetto:1984ik,Catani:2000pi}) that the eikonal factor, $s_{ij}/(s_{iq} s_{jq})$ 
has a physical interpretation in terms of the transverse momentum of the soft gluon, with respect 
to the longitudinal direction singled out by the two hard momenta $p_i$ and $p_j$. This can 
easily be seen by boosting to the center-of-mass frame for $p_i$ and $p_j$, where, up to an 
irrelevant rescaling, we can set
\beq
\label{CMframe}
  \beta_i^\mu \, = \, \{1, 0 , 0, 1 \} \, , \qquad \quad 
  \beta_j^\mu \, = \, \{1, 0 , 0, -1 \} \, , \qquad \quad 
  q^\mu \, = \, \{q^0, {\bf q}_\perp, q^3 \} \, .
\eeq
This leads to 
\beq
\label{qperp}
  \frac{s_{ij}}{s_{iq} s_{jq}} \, = \, \frac{\beta_i \cdot \beta_j}{2 (q \cdot \beta_i)(q \cdot 
  \beta_j)} \, = \, \frac{1}{q^2_{ij}} \, ,
\eeq
where we noted the dependence of the transverse momentum on the chosen dipole and we defined 
$q^2_{ij} \, = \, \left| {\bf q}_\perp \right|^2$ for brevity. We can now rewrite \eq{Cata1},
including the appropriate factor of the strong coupling, as
\beq
\label{Cata2}
  \frac{\alpha_s (\mu^2)}{4 \pi} \,\,  {\bf S}_a^{\lam, (1)} (q) \, = \, {\rm i} C_1 (\e) \,
  \frac{\alpha_s (\mu^2)}{4 \pi} \, \sum_{i\ne j}^nf_{abc} {\bf T}^b_i
  {\bf T}_j^c \left(\frac{\vare^{\lam} \cdot \beta_i}{\beta_i\cdot q}-\frac{\vare^{\lam} 
  \cdot \beta_j}{\beta_j\cdot q}\right) \left(\frac{\mu^2}{q^2_{ij}} \right)^{\! \e} \, ,
\eeq
Next, we observe that a consistent usage of dimensional regularization, when computing a divergent 
quantity such as \eq{Cata2}, requires that one works with the $d$-dimensional version of the running 
coupling, as was the case for virtual corrections to the soft factor (see Refs.~\cite{Magnea:1990zb,
Magnea:2000ss} for a detailed discussion). The $d$-dimensional coupling $\alpha_s(\lambda, \e)$ 
satisfies \eq{beta}, which we rewrite here for easier inspection:
\beq
\label{Beta}
  \lambda \frac{\partial \alpha_s}{\partial \lambda} \, \equiv \, \beta(\alpha_s, \e) \, = \, 
  - 2 \e \alpha_s - \frac{\alpha_s^2}{2 \pi} \, \sum_{k = 0}^\infty \bigg( \frac{\alpha_s}{\pi}
  \bigg)^{\! k} b_k \, .
\eeq
The perturbative solution of this RG equation will be discussed in further detail in the next 
subsection: here, we only need the lowest order approximation, since the the color structure of
\eq{Cata2} first appears at this order. Thus we can use
\beq
\label{RunCouTree}
  \alpha_s (\mu_2^2, \e) \, = \, \left( \frac{\mu_1^2}{\mu_2^2} \right)^{\! \e} \alpha_s (\mu_1^2, \e) 
  \, + \, {\cal O} \left( \alpha_s^2 \right) \, ,
\eeq
reflecting the fact that in $d = 4 - 2 \e$ the gauge coupling is dimensionful, and thus scales 
with its mass dimension. Since $\e < 0$ to regularize IR divergences, \eq{RunCouTree} also reflects
the fact that $\alpha_s (0, \e <0) = 0$, {\it i.e.} the coupling is IR-free at this order, while 
the asymptotically free UV fixed point appears starting at ${\cal O} \left( \alpha_s^2 \right)$.

We now recognize that the scale ratios that appear in \eq{Cata2} raised to the power $\e$ are simply 
responsible for setting the scale of the running coupling to the appropriate transverse momentum, for 
each radiating dipole, a well-established expectation in perturbative QCD~\cite{Bassetto:1984ik}.
\eq{Cata2} becomes
\beq
\label{Cata3}
  \frac{\alpha_s (\mu^2)}{4 \pi} \,\,  {\bf S}_a^{\lam, (1)} (q) \, = \, {\rm i} C_1 (\e) \,
  \, \sum_{i\ne j}^n f_{abc} {\bf T}^b_i {\bf T}_j^c \left( \frac{\vare^{\lam} \cdot 
  \beta_i}{\beta_i\cdot q} - \frac{\vare^{\lam} \cdot \beta_j}{\beta_j \cdot q}\right) \,
  \frac{\alpha_s ( q^2_{ij}, \e )}{4 \pi} \, .
\eeq
Correspondingly, \eq{CelCurr1lo} becomes
\beq
\label{CelCurr1lo2}
  \frac{\alpha_s (\mu^2)}{4 \pi} \, \omega_q \, {\bf S}_a^{+, (1)} \, = \, 
  {\rm i} C_1(\e) f_{abc}  \sum_{i \ne j}^n {\bf T}^b_i {\bf T}_j^c \,\, \frac{z_{ij}}{z_{qi}z_{qj}} 
  \, \frac{\alpha_s ( q^2_{ij}, \e )}{4 \pi} \, ,
\eeq
which is perhaps the most elegant and transparent expression for the one-loop soft current.
Similarly to the case of virtual corrections, each term in the dipole sum in \eq{CelCurr1lo2} 
is the product of a scale-invariant, `celestial' factor, completely dictated by symmetry, times
a coupling factor, where all scale and regulator dependences are confined.

The structure of \eq{CelCurr1lo2} elucidates the origin of all the logarithms in \eq{CelCurr1loExp}:
they are UV logarithms resummed by the renormalization group. This naturally leads to question whether
these logarithms could emerge from a CCFT, or whether they should be treated as `external data', emerging
from the bulk theory, in the same way as the factor $K(\alpha_s, \e)$ in \eq{Suda}: indeed, the two 
are related to each other by unitarity, via the necessary cancellation of infrared poles dictated 
by the KLN theorem. In order to gain further insight into this question, we now examine the information
available on the higher-order structure of the single-emission soft current.


\subsection{The single-emission current at higher orders}
\label{Highloone}

In this section, we discuss the remarkable high-order results that have been obtained in the past few 
years for the single soft current in the context of perturbative QCD, we translate them to the celestial 
language, and we discuss to what extent the considerations of \secn{Oloone} extend beyond one loop. 
Specifically, we note that two-loop corrections to the current are known for amplitudes with any number 
of hard colored particles~\cite{Dixon:2019lnw}, and three-loop corrections have been computed for the 
case of two hard colored particles~\cite{Herzog:2023sgb}. We will mostly concentrate on two-loop 
corrections, whose general structure provides useful insights for the celestial program. Two-leg
three-loop correction, while lacking the color complexity of the full result, provide interesting 
tests of the all-order structure.

At two loops, for the first time, the soft current involves color correlation beyond dipoles. At this
order, as discussed in Ref.~\cite{Dixon:2019lnw}, the current can be arranged as
\beq
\label{ArrTwo}
  {\bf S}^{\lam, (2)}_a (q) \, = \, {\bf S}^{\lam, (2)}_{a, {\rm dip}} (q) \, + \, 
  {\bf S}^{\lam, (2)}_{a, {\rm trip}} (q) \, ,
\eeq
where the two contributions, written directly in celestial coordinates, are given by
\beq
\label{Dix1}
  \omega_q \, {\bf S}^{+, (2)}_{a, {\rm dip}} (q) & = & {\rm i} C_2(\epsilon) f_{abc}
  \sum_{i \ne j} \mathbf{T}_i^{b} \mathbf{T}_j^{c} \, \mathcal{Z}_{ij}^q 
  \left(\Omega_q \left| \mathcal{Z}_{ij}^q \right|^2 \right)^{2 \e} \nonumber \\
  \omega_q \, {\bf S}^{+, (2)}_{a, {\rm trip}} (q) & = & f^{aeb} f^{bcd} 
  \sum_{i \ne k \ne j}\mathbf{T}_i^{c} \mathbf{T}_j^{d} \mathbf{T}_k^{e}
  \left(\Omega_q \left|\mathcal{Z}_{ij}^q \right|^2 \right)^{2\e} \nonumber \\
  && \hspace{1cm} \times \, \biggl(1 + \epsilon^2 \zeta_2 \biggr) 
  \biggl[ \mathcal{Z}_{ik}^q F(z_{iqjk}, \epsilon) - 
  \mathcal{Z}_{jk}^q F(z_{ikjq}, \epsilon) \biggr] \, ,
\eeq
while the negative helicity current is obtained by the usual prescription of swapping barred and unbarred 
variables. Let us briefly consider each color structure in turn.

The form of the dipole contribution in \eq{Dix1} is a straightforward generalization of the one-loop 
result: the only differences are that the dipole transverse momenta are now raised to the power $2 \e$
(an important feature that we will further discuss below), and the two-loop constant $C_2(\e)$ contains
up to fourth-order poles, in keeping with the standard counting of soft and collinear divergences.
Specifically, in QCD one finds~\cite{Dixon:2019lnw}
\beq
\label{C2}
    C_2(\epsilon) & = & \frac{C_A}{2 \epsilon^4} - \frac{b_0}{4\epsilon^3} - 
    \frac{1}{2 \epsilon^2} \Big( \widehat{\gamma}_K^{(2)} - C_A \zeta_2 \Big) 
    + \frac{1}{\epsilon} \bigg[ \frac{19}{54} n_f - C_A \bigg( \frac{11 \zeta_3}{6} + 
    \frac{193}{54} \bigg) - \frac{\zeta_2 b_0}{4} \bigg] \nonumber \\
    & + & C_A \bigg( \frac{7 \zeta_4}{8} - \frac{67 \zeta_2}{36} - \frac{571}{81} \bigg) +
    n_f \bigg( \frac{5 \zeta_2}{18} + \frac{65}{162} \bigg) - 
    \frac{31 \zeta_3 b_0}{6} + \mathcal{O}(\epsilon) \, ,
\eeq
where $\widehat{\gamma}_K^{(2)}$ is the two-loop coefficient of the function in \eq{Cascal}, expanded 
in powers of $\alpha_s/\pi$,
\beq 
\label{CuspTwo}
  \widehat{\gamma}_K^{(2)} \, = \, \bigg( \frac{67}{18} - \zeta(2) \bigg) C_A  - \frac{5}{9} n_f \, .
\eeq 
The tripole contribution to \eq{Dix1} is naturally written in terms of celestial coordinates: indeed, 
the variables $z_{ijkl}$ are the celestial counterparts of the conformal cross-ratios defined in \eq{Cicr}, 
\beq 
\label{CelCicr}
  z_{ijkl} \, = \, \frac{z_{ij} z_{kl}}{z_{ik} z_{jl}}  \qquad \longrightarrow \qquad 
  \rho_{ijkl} \, = \, \left| z_{ijkl} \right|^2 \, ,
\eeq 
while the function $F$ is given by a relatively simple uniform-weight combination of single-valued 
harmonic polylogarithms (SVHPL)~\cite{Brown:2004ugm}, given by
\beq
\label{SVHPL4}
  F(z, \e) & = & \frac{1}{\epsilon^2} \mathcal{L}_0 (z) \mathcal{L}_1 (z) + 
  \frac{1}{3 \epsilon} \Big( \mathcal{L}_0^2(z) \mathcal{L}_1 (z) - 
  2 \mathcal{L}_0 (z) \mathcal{L}^2_1 (z) \Big)
  \nonumber \\
  && - \, \mathcal{L}_1 (z) \left(\frac{2}{9} \mathcal{L}^3_0 (z) + 
  \frac{1}{3} \mathcal{L}_0^2 (z) \mathcal{L}_1 (z) 
  + \frac{13}{18} \mathcal{L}_0 (z) \mathcal{L}_1^2 (z) + 
  \frac{7}{12}\mathcal{L}_1^3 (z) \right) \nonumber \\
  && + \, 2 \mathcal{L}_{1010} (z) + \frac{4}{3} \Big[ 2 \left( \mathcal{L}_{0001} (z) + 
  \mathcal{L}_{0010} (z) \right)+\mathcal{L}_{0011} (z) - \mathcal{L}_{0111} (z) - 
  \mathcal{L}_{1011} (z) - \mathcal{L}_{1100} (z) \Big] \nonumber \\
  && + \, 2 \zeta_2 \Big( 2 \mathcal{L}_{01} (z) - \mathcal{L}_{0} (z) \mathcal{L}_{1} (z) \Big) +
  \frac{40}{3} \zeta_3 \mathcal{L}_{1} (z) + \mathcal{O}(\e) \, .
\eeq 
Recall that, in the conventional counting, SVHPLs with $p$ indices have weight $p$, as does $\zeta(p)$, 
as well as a pole $\e^{-p}$: thus, the expression in \eq{SVHPL4} has uniform weight $w = 4$. 

The appearance of this class of functions is not surprising, and indeed it is a necessary ingredient 
for the consistency of the soft effective theory. In fact, SVHPLs of weight $w = 5$, accompanied by 
a simple pole in $\e$, appear also in the three-loop corrections to the soft anomalous dimension 
matrix $\Gamma_n$, and thus in the three-loop soft factor ${\cal Z}_n$, as indicated in \eq{Delta3},
and furthermore they appear with a matching color structure. This is necessary, since the phase-space 
integral of the square of the soft current must contribute to the cancellation of the soft poles 
due to virtual exchanges. In this case, interfering the two-loop tripole contribution to the 
soft current in \eq{Dix1} with its tree-level counterpart, and integrating over the soft gluon 
phase space, will yield a soft pole accompanied by a combinations of SVHPLs of the appropriate 
weight, as required. We also note that multiple polylogarithms (the more general class of functions 
that includes SVHPLs) are related to iterated integrals on the moduli space of the Riemann sphere 
with $n$ marked points~\cite{Brown:2009qja,Almelid:2017qju}, $\mathcal{M}_{0,n}$, while the subset 
of multiple polylogarithms that only have zero and one as symbol letters is given by harmonic 
polylogarithms~\cite{Remiddi:1999ew}. Such functions appear naturally in the context of two-dimensional 
CFTs, for example in the expansion of conformal blocks. Note also that the arguments of the SVHPLs 
appearing in \eq{Delta3} and in \eq{SVHPL4} are energy independent, contrary to the logarithms 
discussed in \secn{Oloone}. Thus, even if, for the moment, we cannot predict the exact mechanism 
by which this class of functions would arise in the putative CCFT, their emergence is to some 
extent natural, and the celestial sphere is the ideal arena to build such functions from scratch.
A fundamental challenge that any CCFT must face is to generate this class of functions dynamically.

Let us now turn our attention to the energy logarithms emerging from the factors raised to powers
of $\e$ in \eq{Dix1}. Crucially, both the dipole and the tripole contribution have precisely the
$\e$ dependence required to be re-expressed in terms of the $d$-dimensional coupling evaluated
at the scale given by a dipole transverse momentum. To understand this point more precisely,
let's go back to \eq{Beta}, and solve it perturbatively beyond the tree-level accuracy achieved 
by \eq{RunCouTree}. The resulting scale evolution of the $d$-dimensional coupling is given by
\beq
\label{RunCou}
  \alpha_s (\mu_2, \e) \, = \, A_1 \, \alpha_s (\mu_1, \e) + A_2 \, \alpha_s^2 (\mu_1, \e) +
  A_3 \, \alpha_s^3 (\mu_1, \e) \, + \, \mathcal{O} (\alpha_s^4) \, ,
\eeq
where the coefficients $A_i$ depend on the ratio $r_{12} \equiv \mu_1^2/\mu_2^2$, on $\e$, and on 
the coefficient of the $\beta$ function. One finds
\beq
\label{ScaChan}
    A_1 \, = \, r_{12}^\e \, , \quad \, A_2 \, = \, \frac{b_0}{4\pi\e} \, r_{12}^\e 
    \big( r_{12}^\e - 1 \big) \, , \quad \, A_3 \, = \, r_{12}^\e \bigg[ \frac{b_1}{8\pi^2\e} \, 
    \big(r_{12}^{2 \e} - 1 \big) + 
    \left(\frac{b_0}{4\pi\e}\right)^2 \big( r_{12}^\e - 1 \big)^2 \bigg] \, , \qquad
\eeq
which can be easily generalized to higher orders. There are two crucial things to note at this 
point. First, each coefficient in \eq{ScaChan} is finite as $\e \to 0$, as it must: the $d$-dimensional
coupling is a finite function of $\e$, which must reduce to ordinary running coupling when $d = 4$.
Second, \eq{RunCou} applies to the {\it renormalized} coupling, whereas the results we have 
presented for the soft current are {\it bare} quantities. On the other hand, all terms in the 
coefficients $A_k$ in \eq{ScaChan}, and beyond, which are proportional to $r_{12}^{\, p \e}$ with 
$p<k$ arise in explicit calculations from UV counterterms in a minimal scheme: indeed, each counterterm
is built replacing a loop in a Feynman diagram with a pure pole in $\e$, without any associated
scale factor raised to powers of $\e$. Thus, the only terms arising from the change of scale
that contribute to the bare calculation at $k$ loops are those proportional to $r^{k \e}$. It 
is therefore legitimate to take the bare results and upgrade them to the renormalized calculation,
with the understanding that the required UV counterterms must still be added to get the complete 
answer. This is what we did at one loop, leading to \eq{CelCurr1lo2}.

Clearly, changing the scale at which the coupling is evaluated will affect the coefficients of
higher-order terms, reshuffling contributions between different orders. It is straightforward
to evaluate the consequences of this reshuffling: for example, in our case, we can write
\beq
\label{CurrExpRun}
  \omega_q \, \mathbf{S}^{\lam} (q) & = & g_s \omega_q \bigg[ {\bf S}^{\lam, (0)} (q) +
  \alpha_s(\mu_2) \, {\bf S}^{\lam, (1)} (q) + \alpha_s^2 (\mu_2) \, {\bf S}^{\lam, (2)} (q)
  + {\cal O} \left( \alpha_s^3 \right) \bigg] \\
    & = & g_s \omega_q \bigg[ {\bf S}^{\lam, (0)} (q) + \alpha_s (\mu_1) A_1 \, 
    {\bf S}^{\lam, (1)} (q) + \alpha_s^2 (\mu_1) \left( A_1^2 \, {\bf S}^{\lam, (2)} (q) +
    A_2 \, {\bf S}^{\lam, (1)} (q) \right) + \mathcal{O} \left( \alpha_s^3 \right) \bigg] \, .
    \nonumber 
\eeq
Taking this shift into account, and using \eq{CelCurr1lo2} for the one-loop result, at two
loops we get
\beq
\label{Dix2}
  \bigg( \frac{\alpha_s (\mu^2)}{4 \pi} \bigg)^2 \omega_q \, 
  {\bf S}^{+, (2)}_{a, {\rm dip}} (q) \quad & \longrightarrow & \quad {\rm i} \widetilde{C}_2(\e) 
  f_{abc} \sum_{i \ne j} \mathbf{T}_i^{b} \mathbf{T}_j^{c} \, \frac{z_{ij}}{z_{qi} z_{qj}} 
  \bigg( \frac{\alpha_s ( q_{ij}^2 )}{4 \pi} \bigg)^2 \nonumber \\
  \bigg( \frac{\alpha_s (\mu^2)}{4 \pi} \bigg)^2 \omega_q \, 
  {\bf S}^{+, (2)}_{a, {\rm trip}} (q) \quad & \longrightarrow & \quad 
  \Big( 1 + \epsilon^2 \zeta_2 \Big) \, f^{aeb} f^{bcd} 
  \sum_{i \ne k \ne j} \mathbf{T}_i^{c} \mathbf{T}_j^{d} \mathbf{T}_k^{e} \\
  && \hspace{1cm} \times \,  
  \biggl[ \frac{z_{ik}}{z_{qi} z_{qk}} F(z_{iqjk}, \epsilon) - 
  \frac{z_{jk}}{z_{qj} z_{qk}} F(z_{ikjq}, \epsilon) \biggr] 
  \bigg( \frac{\alpha_s ( q_{ij}^2 )}{4 \pi} \bigg)^2 \, , \quad \nonumber 
\eeq
where the shifted prefactor $\widetilde{C}_2(\epsilon)$, in QCD, is given by
\beq
\label{Ctil2}
  \widetilde{C}_2 (\e) & = & C_2 (\e) - \frac{b_0}{\e} C_1(\e) \nonumber \\
  & = & \frac{C_A}{2 \e^4} + \frac{3 b_0}{4 \e^3} 
  - \frac{1}{2 \e^2} \Big( \gamma^{(2)}_K - C_A \zeta_2 \Big)
  + \frac{1}{\e} \bigg[ \frac{19}{54} n_f - 
  \bigg( \frac{11 \zeta_3}{6} +\frac{193}{54} \bigg) C_A  + \frac{\zeta_2}{4} b_0 \bigg] \nonumber\\
  && - \, \bigg( \frac{3 \zeta_2^2}{20} + \frac{571}{81} \bigg) C_A + 
  \frac{65}{162} n_f - \frac{\zeta_2}{2}\gamma^{(2)}_K - \frac{45 \zeta_3}{6} b_0 +
  \mathcal{O}(\e) \, .
\eeq
It is easy to argue that the mechanism described in this section generalizes to all orders in 
perturbation theory and to the multipole color structures that arise beyond two loops. This is
discussed in Ref.~\cite{Herzog:2023sgb}, and the basic argument is very simple: in the bare
current, every added loop is accompanied by an extra factor of $\mu^{2 \e}$; since the only 
other dimensionful quantity in the single soft current is the soft momentum $q^\mu$, dimensional
counting requires a quantity with two negative powers of $q$, also raised to power $\e$; rescaling 
invariance under $\beta_i \to \kappa_i \beta_i$ then uniquely selects the set of $q_{ij}^2$'s as
the only possible such factors for massless emitters. It must therefore be possible to organize
the multi-loop answer for the bare current in terms of a sum over pairs of hard particles,
and the expansion parameter automatically becomes the running coupling evaluated at the scale
of the appropriate transverse momentum.

We conclude that the known structure of the single soft current, to all orders in perturbation 
theory, displays promising features for a connection to a putative CCFT. All contributions to the
current are functions of scale-invariant combinations of distances on the celestial sphere (either 
eikonal ratios such as ${\cal Z}_{ij}^q$, or conformal cross ratios such as $z_{ijkq}$), while
the scale and regulator dependence (up to overall constants) is confined to the running coupling, 
which must be evaluated at scales set by the transverse momenta relative to all possible dipoles. 
Furthermore, the class of functions contributing to multipole corrections at high orders is naturally 
defined on the celestial sphere. {\it All logarithms} arising from the $\e$-expansion of finite 
order results are UV in nature and are reabsorbed in the strong coupling by implementing RG running. 
One could conclude that the CCFT does not need to be logarithmic\footnote{Further evidence in this 
direction is discussed in Ref.~\cite{Agrawal:2025bsy}. We note however that in the literature there 
are arguments in favor of the logarithmic nature of the celestial CFT, beyond the appearance of 
logarithmic branches at loop order: for example, the vanishing of the central charge, and the 
existence of a primary field of dimension 2 \cite{Fiorucci:2023lpb}.}, and scale-setting should be 
treated as `external information' for the CCFT, as is the case for virtual corrections. The alternative
- that the scale logarithms can be generated by a logarithmic CCFT - appears unlikely, but of
course would be remarkable from the point of view of holography, providing an explicit example 
in which bulk scale evolution emerges from a boundary conformal theory.


\section{The double-emission current}
\label{DoubleSoft}

Also in the case of double soft emission, results from the QCD community for massless non-abelian 
amplitudes at tree-level have been available for many years~\cite{Catani:1999ss}. More recently, 
they have been re-derived and generalized to different spins in the context of celestial holography 
and asymptotic symmetries~\cite{Volovich:2015yoa,Klose:2015xoa}. Furthermore, remarkably, the double soft 
current in QCD has recently been computed to one-loop accuracy in Refs.~\cite{Zhu:2020ftr, Czakon:2022dwk}. 
In the context of phenomenology, the primary interest for studying the double soft limit is the 
understanding of amplitude-level factorization, and the application to infrared subtraction for 
collider observables (see, for example,~\cite{Agarwal:2021ais}); on the celestial side, on the other 
hand, double radiation provides an arena where to study a number of interesting limits which
could be controlled by the boundary theory. 

First and foremost, double radiation allows to study the collinear limit, where the two soft gluons 
are emitted in the same direction: this limit should be controlled by the operator product expansion 
(OPE) of the CCFT. In this regard, we emphasize here that an important issue is the order in which 
soft and collinear limits are taken. In the literature, most investigations of the OPE have started 
from the collinear limit of hard amplitudes \cite{Pate:2019lpp,Guevara:2021abz,Himwich:2021dau}: these 
factorize in terms of amplitude-level {\it splitting functions}, which have been studied in QCD for 
many years (see, for example,~\cite{Mangano:1990by,Bern:1993qk,Bern:1994zx}). One may then proceed 
to take a Mellin transform to the celestial amplitude basis, study the OPE of hard operators on the 
celestial sphere and, eventually, take the conformally soft limit, defined by taking the Mellin variable 
$\Delta\rightarrow 1$, in order to study the OPE of soft currents. 

Our viewpoint is different: we start from the assumption that the celestial theory must describe the 
long-wavelength, soft limit of scattering amplitudes: therefore, we always take the soft limit first,
leading to the study of the infrared factor ${\cal Z}_n$ for virtual corrections, and to the study
of soft currents in the case of real radiation. In essence, we always consider the behavior of scattering
amplitudes at {\it leading power} in the soft limit for a set of virtual or real momenta. Only at this 
stage we explore collinear limits, which should be described by a celestial OPE. In this regard, note that
the two order of limits in general {\it do not commute}: for example, the factorization formula in
\eq{factamplvirt}, which is the starting point for the analysis of Ref.~\cite{Magnea:2021fvy}, is derived
under the assumption of {\it fixed-angle} scattering, {\it i.e.} assuming that all Mandelstam invariants
are parametrically of the same order, which excludes collinear and Regge limits. This means for example 
that collinear and Regge logarithms that are not accompanied by infrared singularities are not controlled
by \eq{factamplvirt} (for a more detailed discussion of this point, see for example~\cite{Bret:2011xm,
DelDuca:2011ae}).

One advantage of this way of proceeding is that, order by order in perturbation theory, we can start from
exact results for the soft currents, and these allow in principle to look at the full OPE, complete of all 
non-singular terms. This is not the case if using splitting functions, which retain only singular contributions
in the collinear limit. In other words, we believe that using multiple soft currents potentially opens the way 
to study different questions than those accessible starting from splitting functions. We also note that, when
discussing collinear configurations, we always consider only ``true" collinear limits, defined in momentum space
by taking $q_i^\mu \propto q_j^\mu$ for the momenta of some massless particles $i$ and $j$, as opposed to
``holomorphic'' collinear limits, obtained by taking $z_i \to z_j$ while considering $\bar{z}_i$ and $\bar{z}_j$
as independent.

Finally, we note that multiple soft currents in the context of factorization are defined by taking {\it uniform}
or ``democratic'' soft limits, {\it i.e.} by scaling all soft momenta by the same parameter $\lambda$, as
$q_i^\mu \to \lambda q_i^\mu$, and then taking the leading-power contribution as $\lambda \to 0$ in the
scattering amplitude (the power is $\lambda^{-p}$ when $p$ gluons are taken soft). This is the most general
multiple soft limit, and the one which is relevant for phenomenological applications, such as infrared 
subtraction. Given the uniform soft limit, one can extract from it a range of {\it hierarchical} soft limits,
when subsets of soft particles have energies that are taken to vanish at different rates. Hierarchical, or
``strongly ordered'' soft limits are significantly simpler, although still quite non trivial in non-abelian
theories: they have been studied sistematically in Ref.~\cite{Magnea:2024jqg}, and in detail at one loop for
double currents in Refs.~\cite{Zhu:2020ftr,Czakon:2022dwk}. The relative simplicity of strongly-ordered limits
can provide further hints on the structure of the boundary theory that could underpin them.

We now proceed to discuss gauge-theory results for double soft currents, first at tree level, and then briefly
at one loop, translating these results in celestial coordinates. While our setup is different, as discussed 
above, the results have substantial overlaps with existing literature on multiple soft limits in a celestial 
context~\cite{He:2015zea,Volovich:2015yoa,Bhardwaj:2022anh,Krishna:2023ukw,Bhardwaj:2024wld}, as we will 
note below.


\subsection{The double-emission current at tree level}
\label{Treedouble}

Our natural starting point is the tree-level expression derived in Ref.~\cite{Catani:1999ss}. It is
\beq
\label{DoubCurrCat}
  {\bf J}_{a_1 a_2}^{\mu_1 \mu_2, (0)} (q_1, q_2) & = & \sum_{l \ne m} \frac{p_m^{\mu_1}}{p_m \cdot q_1} 
  \frac{p_l^{\mu_2}}{p_l \cdot q_2} \, {\bf T}_m^{a_1} {\bf T}_l^{a_2} \, + \, 
  \sum_{m} \Bigg[ \frac{p_m^{\mu_1} p_m^{\mu_2}}{p_m \cdot(q_1 + q_2)} 
  \bigg( \frac{{\bf T}_m^{a_2} {\bf T}_m^{a_1}}{p_m \cdot q_2} + 
  \frac{{\bf T}_m^{a_1} {\bf T}_m^{a_2}}{p_m \cdot q_1} \bigg)
  \nonumber \\ 
  && \hspace{1cm} + \, {\rm i} f_{\phantom{a_1 a_2} a}^{a_1 a_2} {\bf T}^a_m \, 
  \frac{p_m \cdot(q_2 - q_1) g^{\mu_1 \mu_2} + 
  2 p_m^{\mu_1}q_1^{\mu_2} - 2 p_m^{\mu_2} q_2^{\mu_1}}{2 (q_1 \cdot q_2)(p_m \cdot(q_1 + q_2))} 
  \Bigg] \, .
\eeq
It is easy to recognize the diagrammatic origin of the various terms in \eq{DoubCurrCat}: the first sum 
arises from independent emissions from two different hard legs, and is essentially the product of two copies
of the single-emission current; the second sum contains double emissions from a single hard leg, with the
term on the second line arising from gluon splitting.

Contracting with the gluon polarization vectors according to \eq{multsgc}, we obtain the eikonal form 
factors for double emission,
\beq
\label{TwoEff}
  {\bf S}_{a_1 a_2}^{\lam_1 \lam_2, (0)} (q_1, q_2) \, = \, \vare_{\mu_1}^{\lam_1} (q_1) \,
  \vare_{\mu_2}^{\lam_2} (q_2) \, {\bf J}_{a_1 a_2}^{\mu_1 \mu_2, (0)} (q_1, q_2) \, .
\eeq
Using the parametrization of momenta and polarizations discussed in \secn{FactRad}, we can recast the results
in celestial coordinates. As was the case for single emission, we multiply the soft factors times the light-cone
energies $\omega_i$ of the emitted gluons, in order to work with dimensionless quantities. After some simple 
manipulations, the result for the emission of two positively polarized gluons can be written as
\beq
\label{Spp}
  \omega_1 \omega_2 \, {\bf S}_{++}^{a_1 a_2, (0)} \, = \, \sum_{l \ne m} \frac{1}{z_{1m}} \frac{1}{z_{2l}} \, 
  {\bf T}_m^{a_1} {\bf T}_l^{a_2} - \frac{1}{z_{12}} \, \sum_m \,
  \frac{z_2 {\bf T}_m^{a_1} {\bf T}_m^{a_2} - z_1 {\bf T}_m^{a_2} {\bf T}_m^{a_1} - {\rm i} z_m 
  f_a^{a_1 a_2} {\bf T}^a_m}{z_{1m} z_{2m}} \, ,
\eeq
or, equivalently, as
\beq
\label{AlSpp}
    \omega_1 \omega_2 \, {\bf S}_{++}^{a_1 a_2, (0)} \, = \, \sum_{l \ne m} 
    \frac{1}{z_{1m}} \frac{1}{z_{2l}} \, {\bf T}_m^{a_1} {\bf T}_l^{a_2} -
    \frac{1}{z_{12}} \sum_m \bigg[ \frac{{\bf T}_m^{a_1} {\bf T}_m^{a_2}}{z_{1m}} - 
    \frac{{\bf T}_m^{a_2} {\bf T}_m^{a_1}}{z_{2m}} \bigg] \, ,
\eeq
where for simplicity we omit the dependence on the soft momenta $q_1$ and $q_2$ on the {\it l.h.s}. 
The expression of the soft factor for the emission of two gluons with different polarizations is 
significantly more intricate and less appealing, involving a residual dependence on ratios of soft-gluon 
energies. It can be written as
\beq 
\label{Spm}
  \omega_1 \omega_2 \, {\bf S}_{+-}^{a_1 a_2, (0)} & = & \sum_{l \ne m} \frac{1}{z_{1m}} \frac{1}{\zbar_{2l}} \, 
  {\bf T}_m^{a_1} {\bf T}_l^{a_2} \, + \, \sum_{m} \frac{1}{\omega_1 |z_{1m}|^2 + \omega_2 |z_{2m}|^2} \,
  \bigg[ \frac{\omega_1 \zbar_{1m} {\bf T}_m^{a_2} {\bf T}_m^{a_1}}{\zbar_{2m}} + 
  \frac{\omega_2 z_{2m} {\bf T}_m^{a_1} {\bf T}_m^{a_2}}{z_{1m}} \nonumber \\
  && - \,  
  {\rm i} f_{\phantom{a_1 a_2} a}^{a_1 a_2} {\bf T}^a_m 
  \left( \frac{\omega_1 \zbar_{1m}}{\zbar_{12}} - 
  \frac{(\omega_1 |z_{1m}|^2 - \omega_2 |z_{2m}|^2)}{2 |z_{12}|^2} + 
  \frac{\omega_2 z_{2m}}{z_{12}} \right) \bigg] \, .
\eeq
We emphasize that \eq{Spp} and \eq{Spm} are exact, providing the complete tree-level result for the uniform
double soft limit, without any collinear or strongly-ordered approximation. The other two possible cases,
$\mathcal{S}_{--}^{a_1 a_2, (0)}$ and $\mathcal{S}_{-+}^{a_1 a_2, (0)}$, are analogous and can be obtained 
by suitably swapping barred and non-barred variables.

It is immediately evident that, from the point of view of celestial holography, \eq{Spp} is rather appealing:
first of all, the double current is holomorphic (and its negatively helicity counterpart is anti-holomorphic);
furthermore, as seen in \eq{AlSpp}, all terms are characterized by pairs of collinear poles, in each case with 
no finite remainders, which suggests a simple behavior under OPE. \eq{Spm}, on the other hand, is remarkably
different: energy dependence does not cancel, so that the final results still depends on the dimensionless
ratios $x_{12} \equiv \omega_1/\omega_2$ and $x_{21} = 1/x_{12}$; furthermore, the expression is neither 
holomorphic nor anti-holomorphic, and its behavior under collinear limits is rather untransparent. In order 
to gain further insights in the content of \eq{Spp} and \eq{Spm}, let us now consider in detail the physically 
relevant limits.

We begin with the physical collinear limit, taking $z_1 \to z_2$, and treating $\bar{z}_i$ as the complex 
conjugate of $z_i$. For the double positive-helicity current, one easily finds
\beq 
\label{CollSpp}
  \omega_1 \omega_2 \, {\bf S}_{++}^{a_1 a_2, (0)} \, \longrightarrow \,\, - \, 
  \frac{{\rm i} f^{a_1 a_2}_{\phantom{a_1 a_2} a}}{z_{12}} \, \sum_m \frac{{\bf T}^a_m}{z_{1m}}
  \, = \, - \, \frac{{\rm i} f^{a_1 a_2}_{\phantom{a_1 a_2} a}}{z_{12}} 
  \, \omega_1 \, {\bf S}_{+}^{a, (0),} \, .
\eeq 
up to corrections regular as $z_1 \to z_2$. Turning to the case of mixed helicity currents, one finds 
\beq
\label{CollSpm}
  \omega_1 \omega_2 \, {\bf S}_{+-}^{a_1 a_2, (0)} \, \longrightarrow \,\, - \, 
  \frac{{\rm i} f^{a_1 a_2}_{\phantom{a_1 a_2} a}}{\zbar_{12}} \, \frac{1}{1 + x_{21}} \, 
  \sum_m \frac{{\bf T}^a_m}{z_{1m}} \, - \, \frac{{\rm i} f^{a_1 a_2}_{\phantom{a_1 a_2} a}}{z_{12}} \,
  \frac{1}{1 + x_{12}} \, \sum_m \frac{{\bf T}^a_m}{\zbar_{1m}} \, ,
\eeq 
again up to corrections regular as $z_1 \to z_2$ and $\bar{z}_1 \to \bar{z}_2$. Both these expressions
are simpler and more transparent than the full results: interestingly, but not surprisingly, the collinear
limit in both cases selects the explictly non-abelian terms in the double current. The reason, from the
viewpoint of QCD, is that dominant diagrams in the collinear limit are those when the two final-state
gluons emerge from a single off-shell gluon, which goes to its mass shell when the two emitted gluons 
become collinear. More specifically, while \eq{CollSpp} begins to bear a resemblance to a current algebra
of Ka\v c-Moody type, it is important to notice that the {\it l.h.s.} of \eq{CollSpp} is {\it not} the
product of two single-particle soft currents: in the non-abelian theory, double emission is always
correlated, and indeed the color correlation is all that survives in the collinear limit.

To further clarify this point, we now briefly consider the {\it strongly-ordered} soft limit in which
the gluon with momentum $q_2$ is much softer than the gluon with momentum $q_1$, {\it i.e.} $\omega_2 
\ll \omega_1$. In this limit \eq{DoubCurrCat} becomes~\cite{Catani:1999ss}
\beq 
\label{SOCat}
  {\bf J}_{a_1 a_2}^{\mu_1 \mu_2, (0)} (q_1, q_2) \Big|_{\rm s.o.} \, = \, 
  \left( \delta_{a_1}^{\phantom{a} a} \, {\bf J}_{a_2}^{\mu_2, (0)} (q_2) + 
  {\rm i} f_{a_1 a_2}^{\phantom{a_1 a} a} \, \frac{q_1^{\mu_2}}{q_1 \cdot q_2} \right) 
  {\bf J}_{a}^{\mu_1, (0)} (q_1) \, ,
\eeq
which, in celestial coordinates, yields
\beq
\label{SOSp}
  \omega_1 \omega_2 \, {\bf S}_{++}^{a_1 a_2, (0)} \Big|_{\rm s.o.} & = &  
  \left( \delta^{a_1}_{\phantom{a} a} \, \omega_2 {\bf S}_+^{a_2, (0)} - {\rm i} \,
  \frac{f^{a_1 a_2}_{\phantom{a_1 a2} a}}{z_{12}} \right) \omega_1 {\bf S}_+^{a, (0)} \, ,
  \nonumber \\ 
  \omega_1 \omega_2 \, {\bf S}_{+-}^{a_1 a_2, (0)} \Big|_{\rm s.o.} & = &  
  \left( \delta^{a_1}_{\phantom{a} a} \, \omega_2 {\bf S}_-^{a_2, (0)} - {\rm i} \,
  \frac{f^{a_1 a_2}_{\phantom{a_1 a2} a}}{\bar{z}_{12}} \right) \omega_1 {\bf S}_+^{a, (0)} \, ,
\eeq
so that the collinear limit of \eq{SOSp} indeed reproduces the strongly-ordered limit of \eq{CollSpp} 
and \eq{CollSpm}, reached by taking $x_{12} \to 0$, or $x_{21} \to 0$. We note in passing that the
re-factorization of multiple soft currents in strongly-ordered limits is well understood in QCD, 
at tree level for any number of emissions, and at loop level insofar as the general structure is 
concerned~\cite{Magnea:2024jqg}. Essentially, when gluon energies are strongly ordered, one can
organize the emissions by having the hardest gluon emitted from the original system of $n$ Wilson 
lines representing the $n$ hard particles, while the softer gluon sees the harder one as an extra
Wilson line, and is thus effectively radiated by a system of $n+1$ Wilson lines.

We now need to ask how the structures that we have just described might arise in the context of
a celestial theory. The most natural guess is that double emission currents should be computed
by correlators involving two copies of the celestial Noether current, generalizing \eq{WId}. For 
example, using two holomorphic currents should lead to an expression of the form
\beq
\label{WId2}
  \Big\langle j^{a_1} (u_1) j^{a_2} (u_2) \prod_{k=1}^n V (z_k, \bar{z}_k) \Big\rangle \, \sim \, 
  \omega_1 \omega_2 \, S_{++}^{a_1 a_2, (0)} (q_1, q_2) \, 
  {\cal C}_n \big( \{ z_i \}, \kappa \big) \, ,
\eeq
where here we denoted by $u_1$ and $u_2$ the celestial coordinates associated with the soft 
momenta $q_1$ and $q_2$. We note that \eq{AlSpp} is rather promising in this respect: the soft
factor is a sum of terms that would naturally arise from the OPE of each current with any one
of the vertex operators, plus a term that would arise from the OPE of the two holomorphic 
currents. Furthermore, \eq{CollSpp} would constrain the OPE of the holomorphic currents
to be of the form
\beq 
\label{OPE++}
    j^{a_1} (u_1) \, j^{a_2} (u_2) \, \sim \, {\rm i} \, 
    \frac{f^{a_1 a_2 a}}{u_{12}} \, j_a (u_2) \, ,
\eeq 
which is the structure expected in CFT in the presence of a Ka\v c-Moody symmetry, as noted already 
in Ref.~\cite{He:2015zea}. One easily sees that anti-holomorphic currents should provide a second
copy of this Ka\v c-Moody OPE, reflecting the form of the current for the emission of two negative-helicity
gluons. This interpretation is however severely challenged by the mixed-helicity soft factor in 
\eq{Spm}, leading to the collinear limit in \eq{CollSpm}. This would suggest for the OPE of a holomorphic 
with an anti-holomorphic current the form
\beq 
\label{OPE+-}
    j^{a_1} (u_1) \, \bar{j}^{a_2} (\bar{u}_2) \, \sim \,
    {\rm i} \frac{f^{a_1 a_2 a}}{(1 + x_{21}) \zbar_{12} } j_a(u_1) + 
    {\rm i} \frac{f^{a_1 a_2 a}}{(1 + x_{12}) z_{12}} \bar{j}_a (\bar{u}_2) \, .
\eeq 
We regard \eq{OPE+-} as very challenging for the CCFT perspective, but at the same time very informative.
First of all, note that any CFT supporting holomorphic factorization would require a vanishing {\it r.h.s.}
for \eq{OPE+-}. The bulk gauge theory however requires a non-vanishing result, ruling out factorizable
theories for the CCFT. An example would be the WZNW model, which reduces to the free boson in \eq{freebos}
in the non-interacting limit: note, however, that the WZNW model can be independently ruled out by showing
that the explicit form of the correlator ${\cal C}_n$ computed in the bulk theory does not satisfy the 
appropriate Knizhnik-Zamolodchikov equation~\cite{Nastase:2021izh}. Perhaps even more challenging is the 
fact that the holomorphic and anti-holomorphic factors on the {\it r.h.s.} of \eq{OPE+-} are weighted by
factors which depend on gluon energies in the bulk theory, and have no obvious interpretation on the
celestial sphere\footnote{Note that, defining the energy fractions $y_i \equiv \omega_i/(\omega_1 + 
\omega_2)$, so that $y_1 + y_2 = 1$, one can write $1/(1 + x_{21}) = y_1$, and similarly $1/(1 + x_{12}) 
= y_2$, a fact that will be useful also in \secn{TripleSoft}.}.

A range of tentative solutions have been proposed in the literature to tackle these difficulties, beginning
with Ref.~\cite{He:2015zea}, where a strong ordering for the energies of the soft gluons (say, $x_{21} \to 0$) 
was introduced, in order to discard one of the two terms in \eq{OPE+-}. In such a limit, one can interpret 
one of the two currents as generating a Ka\v c-Moody symmetry, while the second current transforms under 
the action of the first. A subsidiary problem in this approach is the fact that the two possible strong 
orderings do not give the same result in the non-abelian theory: in fact, they give either one or the other 
of the two terms on the {\it r.h.s.} of \eq{OPE+-}. As a consequence, the OPE of the two celestial currents
is not uniquely defined on the soft boundary of phase space\footnote{Recently, a prescription to 
handle this ambiguity was proposed in Ref.~\cite{Pranzetti:2025flv}, using the technology of shadow transforms
in the context of celestial amplitudes, based on earlier work in Ref.~\cite{Freidel:2021ytz}. A formal analysis 
of this problem, linking it to the curvature of the vacuum manifold for Yang-Mills theory, was presented in 
Ref.~\cite{Kapec:2022hih}.}. We would argue that \eq{CollSpm} provides a solution of this problem, dictated 
by bulk gauge theory, in the form of a continuous interpolation between the two strong-ordered limits,
parametrized by the energy ratio $x_{12}$: in other words, there is no ambiguity in \eq{CollSpm}, and therefore 
in \eq{OPE+-}. From the perspective of a QCD practitioner, while it is reasonable to expect a simplified 
behavior in the strongly-ordered limit, there are no reasons to restrict the analysis to this case: the 
expression in Ref.~\cite{Catani:1999ss} is derived without making any assumptions about the helicities 
of the two gluons, and there are certainly cases of interest where one would want to consider gluons 
of different helicities, but with comparable ``softness". With this in mind, we believe one should take 
\eq{OPE+-} at face value, and investigate whether an OPE of this kind can arise from a consistent theory. 
A step in this direction is taken in Section~\ref{TripleSoft}.


\subsection{The double-emission current at higher orders}
\label{Olotwo}

The one-loop correction to the double soft emission current in the general case involving $n$ hard
particles was computed quite recently, in Refs.~\cite{Zhu:2020ftr,Czakon:2022dwk}. It is a challenging
calculation even with modern techniques: indeed, while most required master integrals can be computed
exactly in terms of hypergeometric functions involving $\e$ as a parameter, the full answer is known
only as an expansion in powers of $\e$, starting of course with a double pole, $\e^{-2}$. The resulting
expansion contains both logarithms and polylogarithms (in particular dilogarithms in the finite 
contributions) whose arguments involve various ratios of physical scales. The dimensional argument
used in \secn{SingSoft} to reabsorb all logarithms in the running of the $d$-dimensional  
coupling is not available for double emission: first of all, even for double emission from a single 
hard dipole, one can define several possible transverse momenta; furthermore, with two available 
soft momenta, one can construct scale-invariant combinations with mass dimension (-2) which cannot
be interpreted as transverse momenta, such as $(\beta_i \cdot \beta_j)/(\beta_i \cdot q_1 \beta_j 
\cdot q_2)$. Finding the best scale choices for the organization of the renormalized double soft
current is therefore still an open problem.

Since the full expression for the one-loop double current is very lengthy and, at this stage, rather
untransparent in celestial coordinates, in this section we focus most of our analysis on the strongly-ordered 
limit, referring the reader to the original references for further study. This limit is considerably simpler, 
and allows to catch a glimpse of structures that will appear at even higher order. In particular, taking 
collinear limits in the strongly-ordered configuration, one may conjecture the form of the celestial soft 
OPE, as was done in the previous section, taking into account loop corrections. Considering color structures, 
we note that the current receives contributions both from abelian-like and purely non-abelian terms, exactly 
as was the case at tree-level. The abelian-like term is simply the symmetrized product of a one-loop and a 
tree-level single emission currents, and it becomes subleading (and non-singular) in the collinear limit. 
The non-abelian term, instead, displays the color structures
\beq 
\label{ColStru}
  {\rm i} f_{a_1 a_2 a} \mathbf{T}^a_i \, , \qquad \quad
  f_{a_1 b e} \, f^e_{\phantom{e} a_2 c} \, \mathbf{T}^b_i \mathbf{T}^c_j \, .
\eeq
The second structure, quadratic in the structure constants, like the tripole contribution to the single
soft current at two loops in \eq{Dix1}, is expected, since the interference of the one-loop double
current with its tree-level counterpart must contribute to the cancellation of the infrared poles of
the three-loop quadrupole virtual correction.

Let us now concentrate on the strongly-ordered limit, which will allow us to draw some general consequences 
for the structure of the celestial OPE. The factorization properties of soft currents in strongly-ordered 
limit have been recently studied in the context of subtraction in Ref.~\cite{Magnea:2024jqg}, and the 
general results derived there agree with the one-loop discussion in Ref.~\cite{Zhu:2020ftr}, which we 
partly reproduce here. The basic idea, already mentioned in the previous section, is that, in the case 
of double emission (or indeed multiple emission) with strongly ordered energies, the softest gluon sees 
all other particles in the amplitude, including the second gluon, as hard. In other words, 
at leading power, we can think of the softest gluon as being emitted from a system of $n+1$ Wilson lines, 
with the $(n+1)$-st line representing the harder gluon. This harder gluon, on the other hand, is emitted
by the original system of $n$ Wilson lines. 

With this in mind, it is natural to organize the strongly-ordered double radiation in two terms: first,
the two gluons can be independently emitted by the original $n$ Wilson lines; second, we have to take 
into account the correlated, purely non-abelian emission of the softest gluon from the Wilson line 
associated with the harder one. In formulas, choosing for example $\omega_1 \ll \omega_2$, we can write
\beq 
\label{SOSoftOrg}
  \langle a_1, \lambda_1; a_2, \lambda_2 | \mathcal{M}_n^{\, (gg)} \rangle \, \, 
  \substack{\phantom{some stuff} \\ \longrightarrow \\ {\scriptscriptstyle \omega_1 \ll \omega_2 \to 0}} \, \,
  \Big[ \, {\bf S}^{a_1}_{\lam_1} \, {\bf S}^{a_2}_{\lam_2} + 
  \Delta{\bf S}^{a_1 a_2}_{\lam_1 \phantom{a_1} a} \, {\bf S}^{a}_{\lam_2} \Big] \, 
  | \mathcal{M}_n \rangle \, ,
\eeq 
where the left-hand side represents the projection of the full double-emission matrix element (from $n$ 
hard particles) on a specific color and helicity state for the two soft gluons. On the right-hand side,
the first term is the ordered product of two single soft currents (an abelian-like contribution), while
the second term accounts for the non-abelian emission of the softer gluon from the harder one. Notice
that we argue here that this organization of the strongly-ordered double current applies to all orders.
At this point, we can exploit the factorization properties on the left-hand side of \eq{SOSoftOrg}, namely
\beq 
\label{BracketFac}
  \langle a_1, \lambda_1; a_2, \lambda_2 | \mathcal{M}_n^{\, (gg)} \rangle \, = \, 
  {\bf S}_{\lam_1 \lam_2}^{a_1 a_2} \, |\mathcal{M}_n \rangle \, ,
\eeq 
to write the general result, valid to all orders,
\beq 
\label{SOSoftOrg2}
  {\bf S}_{\lam_1 \lam_2}^{a_1 a_2} \, \, \substack{\phantom{some stuff} \\ \longrightarrow \\ 
  {\scriptscriptstyle \omega_1 \ll \omega_2 \to 0}} \, \,
  {\bf S}^{a_1}_{\lam_1} \, {\bf S}^{a_2}_{\lam_2} + 
  \Delta{\bf S}^{a_1 a_2}_{\lam_1 \phantom{a_1} a} \, {\bf S}^{a}_{\lam_2} \, .
\eeq 
When we take the collinear limit to determine the singular part of the OPE in the celestial version of 
\eq{SOSoftOrg2}, the first, abelian-like term is always sub-leading, because it does not contain terms 
that depend on both gluon momenta in a singular way: both gluons are emitted from the original system 
of $n$ Wilson lines. In the collinear limit, we are thus allowed to focus solely on the second term. 
To obtain the non-abelian factor in \eq{SOSoftOrg2}, representing the emission of the softest gluon 
from the harder one, we need to identify one of the color insertion operators ${\bf T}_i$ in the definition 
of the single soft current as acting on the harder gluon emitted by the last factor in the second term
of \eq{SOSoftOrg}. Acting on a line with an adjoint index, the operator ${\bf T}_i$ evaluates to a 
structure constant: specifically, we need $(T_2^b)_{a_2 a} = {\rm i} f_{a_2 b a}$, where we have 
set $i = 2$.

We can check the validity of our reasoning at lowest order by evaluating \eq{SOSoftOrg2} at tree-level 
and verifying that it matches the known results, \eq{SOSp}. As expected, for example in the case of 
mixed-helicity gluons, we find
\beq 
\label{CheckSOSoft}
  {\bf S}_{+ -}^{a_1 a_2, (0)} \, &
  \substack{\phantom{some stuff} \\ \longrightarrow \\ {\scriptscriptstyle \omega_1 \ll \omega_2 \to 0}} & \,
  {\bf S}^{a_1, (0)}_+ \, {\bf S}^{a_2, (0)}_- \, + \, \Delta{\bf S}^{a_1 a_2, (0)}_{+ \phantom{a_1 a_2} a} \, 
  {\bf S}^{(0), a}_- 
  \nonumber \\
  & = & \bigg[ {\bf S}^{a_1, (0)}_+ \delta_{\phantom{a_2} a}^{a_2} - \frac{{\rm i}}{\omega_1} 
  \frac{f^{a_1 a_2}_{\hspace{18pt} a}}{z_{12}} \bigg] \, {\bf S}^{a, (0)}_- \, ,
\eeq  
matching the second line of \eq{SOSp}. Comforted by the tree-level result, we go on to evaluate \eq{SOSoftOrg2} 
to one-loop order, obtaining
\beq 
\label{StrOrd1lo}
  {\bf S}^{a_1 a_2, (1)}_{\lam_1 \lam _2} \, \, 
  \substack{\phantom{some stuff} \\ \longrightarrow \\ 
  {\scriptscriptstyle \omega_1 \ll \omega_2 \to 0}} \, \, 
  {\bf S}^{a_1, (0)}_{\lam_1} \, {\bf S}^{a_2, (1)}_{\lam _2} + 
  {\bf S}^{a_1, (1)}_{\lam_1} \, {\bf S}^{a_2, (0)}_{\lam _2} + 
  \Delta{\bf S}^{a_1 a_2, (0)}_{\lam_1 \phantom{a_1} a} \, {\bf S}^{a, (1)}_{\lam _2} +
  \Delta{\bf S}^{a_1 a_2, (1)}_{\lam_1 \phantom{a_1} a} \, {\bf S}^{a, (0)}_{\lam_2} \, . \quad 
\eeq 
As before, the collinear limit is dominated by the last two terms in \eq{StrOrd1lo}, which carry the 
non-abelian structure. The first of those is the tree-level non-abelian contribution appearing in 
\eq{CheckSOSoft}, multiplied times the one-loop soft current for the emission of the harder gluon. Thus, 
for example in the same-helicity case,
\beq
\label{FirstTree}
  \Delta{\bf S}^{a_1 a_2, (0)}_{+ \phantom{a_1} a} \, {\bf S}^{a, (1)}_+ \, = \, 
  - \frac{{\rm i}}{\omega_1} \frac{f^{a_1 a_2}_{\hspace{18pt} a}}{z_{12}} \,\,  {\bf S}^{a, (1)}_+ \, .
\eeq
The last term in \eq{StrOrd1lo}, on the other hand, is the non abelian term in the one-loop current 
of the softest gluon, multiplied times the tree-level current of the harder gluon. Using the definition 
of the one-loop current in \eq{Cata1} and in \eq{CelCurr1lo}, we note that relevant scale to be raised 
to the power $(-\e)$ is the transverse momentum of gluon 1 with respect to the dipole $(2j)$, which we
denote by $q^2_{1, 2 j} \equiv (q_1 \cdot p_j \, q_1 \cdot q_2)/(q_2 \cdot p_j)$. We emphasize that,
in the collinear limit $q_1 \parallel q_2$, this becomes independent of $j$. We can then write
\beq
\label{DefDeltaJ1J0}
  \frac{\alpha_s (\mu^2)}{4 \pi} \, \Delta{\bf S}^{a_1 a_2, (1)}_{+ \phantom{a_1} a} \, 
  {\bf S}^{a, (0)}_+ \, = \, \bigg[ - \frac{C_1 (\epsilon)}{\omega_1} \, \sum_{j=1}^n  
  f^{a_1}_{\phantom{a_1} b c} \, f^{a_2 b}_{\phantom{a_2 b} a} \, {\bf T}_j^c \, 
  \mathcal{Z}^1_{2j} \, \frac{\alpha_s(q^2_{1, 2 j})}{4\pi} \bigg] 
  \bigg( \frac{1}{\omega_2} \sum_{k=1}^n \frac{{\bf T}_k^a }{z_{2k}} \bigg)
\eeq
where we have again substituted $i \to 2$. This fixes the representation of ${\bf T}_i \to {\bf T}_2$ to 
be the adjoint, picks a particular kinematic factor, $\mathcal{Z}_{2j}^1$, out of the $n+1$ possible choices, 
and fixes the scale at which we must evaluate the coupling constant to $q^2_{1, 2 j}$. A useful way to 
rewrite \eq{DefDeltaJ1J0} is
\beq 
\label{DeltaJ1J0}
  \frac{\alpha_s (\mu^2)}{4 \pi} \, \Delta{\bf S}^{a_1 a_2, (1)}_{+ \phantom{a_1} a} \, 
  {\bf S}^{a, (0)}_+ \, = \, - \frac{C_1 (\e)}{\omega_1 \omega_2} \, \sum_{j \ne k = 1}^n  
  f^{a_1}_{\phantom{a_1} b c} \, f^{a_2 b}_{\phantom{a_2 b} a} \, {\bf T}_j^c \, {\bf T}_k^a \, 
  \mathcal{Z}^1_{2j} \, \frac{\alpha_s(q^2_{1, 2 j})}{4\pi} \,  \bigg( \frac{1}{z_{2k}} - \frac{1}{z_{2j}}
  \bigg) \, , \quad 
\eeq
where we used color conservation to subtract the last term, which gives a vanishing contribution because 
it does not contain any kinematic factor depending on the index $k$. Taking the collinear limit of this 
expression removes any dependence on the index $j$ in the kinematic factor and hence also in the coupling 
constant, since
\beq
\label{CollZ}
  \mathcal{Z}^1_{2j} \, = \, \frac{z_{2j}}{z_{12} z_{1j}}
  \substack{\phantom{some stuff} \\ \rightarrow \\ 
  {\scriptscriptstyle z_1 \to z_2}} \frac{1}{z_{12}} \, .
\eeq
We can now use again color conservation to write the leading term of \eq{DeltaJ1J0} in the collinear 
limit as
\beq 
\label{CollDeltaJ1J0}
  \frac{\alpha_s (\mu^2)}{4 \pi} \, \Delta{\bf S}^{a_1 a_2, (1)}_{+ \phantom{a_1} a} \, 
  {\bf S}^{a, (0)}_+ \!\!\!\!
  & \substack{\phantom{some stuff} \\ \rightarrow \\ {\scriptscriptstyle z_1 \to z_2}} & \!\!\!\!
  \frac{\alpha_s(q^2_{1, 2 j})}{4\pi} \, \frac{C_1(\e)}{\omega_1 \omega_2} \, 
  \frac{1}{z_{12}} \sum_{j \ne k = 1}^n f^{a_1}_{\phantom{a_1} b c} \, 
  f^{a_2 b}_{\phantom{a_2 b} a} \, {\bf T}_j^c \, {\bf T}_k^a \, 
  \bigg(\frac{1}{z_{2j}} - \frac{1}{z_{2k}} \bigg) \nonumber \\
  & = & \!\!\!\!\!\! - \, \frac{\alpha_s(q^2_{1, 2 j})}{4\pi} \, 
  \frac{C_1(\epsilon)}{\omega_1} \, \frac{1}{z_{12}} \,
  \frac{C_A}{2} f^{a_1 a_2}_{\phantom{a_1 a_2} a} \, {\bf S}^{a, (0)}_+ \, ,
\eeq 
where we used the identity
\beq
\label{ColId} 
  - \, f_{a_1 b c} f_{a_2 b a} \sum_{i \ne j} {\bf T}_i^a {\bf T}_j^c \, \big( F(i) - F(j) \big)
  \, = \, \frac{C_A}{2} \, f_{a_1 a_2 a} \sum_i {\bf T}_i^a \, F(i) \, ,
\eeq 
and we note again that $\alpha_s(q^2_{1, 2 j})$ does not depend on $j$ in the collinear limit.
Adding together all non-abelian contributions, the leading term in the collinear (and strongly-ordered) 
limit reads
\beq
\label{CollTot}
  \frac{\alpha_s (\mu^2)}{4\pi} \, {\bf S}^{a_1 a_2, (1)}_{+ +} \!\!\!
  \substack{\phantom{some stuff} \\ \rightarrow \\ {\scriptscriptstyle z_1 \to z2}} \!\!\!
  - \, \frac{1}{\omega_1} \, \frac{f^{a_1 a_2}_{\hspace{18pt} a}}{z_{12}} 
  \bigg[ \, \frac{\alpha_s(q^2_{1, 2 j})}{4\pi} \, \frac{C_A}{2} \, C_1(\epsilon) \, 
  {\bf S}^{a, (0)}_+ + {\rm i} \, {\bf S}^{a, (1)}_+ \bigg] \, .
\eeq
Note that the one-loop OPE in the strongly-ordered limit shows contributions both from the one-loop 
and the tree-level single soft currents. Also in the strongly-ordered limit, it is not easy to envisage 
how \eq{CollTot} could emerge from a Ka\v c-Moody symmetry, even as the simple pole and the color
structure have the correct form: the weights of the two perturbative contributions on the {\it r.h.s.} 
are different, so that the full single soft current is not reconstructed. Our result has the 
same structure as the expressions for the one-loop soft OPE given in Refs.~\cite{Bhardwaj:2024wld,
Krishna:2023ukw}, albeit in a very different notation, reflecting the different approaches that have 
been employed. 

Because \eq{SOSoftOrg2} is an all-order statement, and we have argued that in the collinear limit the 
leading contribution are given by non-abelian factors, it seems reasonable to express the $L-$loop 
celestial strongly-ordered soft OPE as
\beq
\label{OrderL}
  {\bf S}^{a_1 a_2, (L)}_{\lam_1 \lam _2} \, \sim \, 
  \sum_{l = 0}^{L} \Delta{\bf S}^{a_1 a_2, (l)}_{\lam_1 a} \, {\bf S}^{(L - l) a}_{\lam _2} \, .
\eeq
This would mean that, in the strongly-ordered limit, the knowledge of the single soft current at $l$-th 
order would suffice to determine the OPE at that order. This is in stark contrast to what happens away from
the strongly-ordered limit, where the soft current for the double emission contains genuinely new information. 
\eq{OrderL} suggests a rather intricate situation at higher orders, with contributions from all the lower 
orders currents. This perspective challenges the emergence of a Ka\v c-Moody symmetry for the soft currents, 
even in the strongly-ordered limit, when one includes quantum corrections.


\section{The triple-emission current}
\label{TripleSoft}

The last bit of information that we can extract from known contributions on the gauge theory side of 
the celestial correspondence is the case of triple soft emissions. The triple soft current was computed 
at tree-level in Ref.~\cite{Catani:2019nqv}. The study of this quantity in celestial coordinates allows 
us to study the associativity of the OPE of the currents, at tree-level: in fact, in a sensible CFT we 
expect to be able to take multiple OPEs between pairs of currents in succession, and find that the 
result does not depend on the order. We have already seen in \secn{DoubleSoft} that gauge-theory data
pose significant challenges in this direction: for double emission, at leading power, one finds an 
OPE which is unambiguous, as in \eq{OPE++} and in \eq{OPE+-}, however in the mixed helicity case the
putative OPE depends on a continuous parameter related to gluon energy ratios. Taking directly the 
strongly ordered limit removes the parameter dependence, but effectively renders the OPE ambiguous. 
As we will see in this section, a similar problem affects associativity, which can be checked starting 
from the triple soft current. Our discussion in this section is based on gauge-theory data, but our 
results mirror the conclusions of Ref.~\cite{Ball:2022bgg}, where the analysis was framed directly 
in terms of celestial currents. Ref.~\cite{Ball:2022bgg} also proposed a motivation for the failure 
of associativity of the OPE of the soft currents, arguing that the naive formulation of soft insertions 
cannot be interpreted in terms of local operators on the celestial sphere. Subsequent papers have 
explored the consequences of this argument, conjecturing connections with shadow operators (already 
mentioned in~\cite{Kapec:2021eug}, and more recently in Refs.~\cite{Himwich:2025bza, Pranzetti:2025flv}), 
and with multi-particle states on the celestial sphere~\cite{Ball:2023sdz, Ball:2024oqa, Kulp:2024scx}.

Our starting point, on the gauge theory side, is the full expression for the triple soft 
current~\cite{Catani:2019nqv},
\beq 
\label{TripleCat}
  {\bf J}_{a_1 a_2 a_3}^{\mu_1 \mu_2 \mu_3} (q_1, q_2, q_3) & = & \big[ {\bf J}_{a_1}^{\mu_1} (q_1) \,
  {\bf J}_{a_2}^{\mu_2}(q_2) \, {\bf J}_{a_3}^{\mu_3}(q_3) \Big]_{\rm sym} \nonumber \\
    & + & \left[ \left( {\bf J}_{a_1}^{\mu_1} (q_1) \, {\bf \Sigma}_{a_2 a_3}^{\mu_2 \mu_3}(q_2, q_3) 
    \right)_{\rm sym} + (1 \leftrightarrow 2) + (1 \leftrightarrow 3) \right] \nonumber \\
    & + & {\bf \Gamma}_{a_1 a_2 a_3}^{\mu_1 \mu_2 \mu_3} (q_1, q_2, q_3) \, ,
\eeq 
where the first line is the expected abelian-like contribution, where gluons are attached to different 
hard particles, the second line represents terms with two of the gluons attaching to the same hard 
particle, and, finally, the third line is the contribution from diagrams where all gluons attach to 
the same leg; the subscript ``$\mathrm{sym}$'' means the symmetrized product. The first two lines are 
not relevant for the discussion in this section and are easily understood in terms of the single 
and double currents that we have already presented: for example, ${\bf  \Sigma}_{a_2 a_3}^{\mu_2 
\mu_3}$ is the non-abelian contribution to the double soft current, displayed on the second line 
of \eq{DoubCurrCat}. We focus therefore on the third line in \eq{TripleCat}, describing three 
gluons emitted from the same hard particle: this is the most complex and interesting term, and 
the only one containing singular contributions in triple collinear limits. Explicitly, it reads
\beq 
\label{TripleCat1}
  {\bf \Gamma}_{a_1 a_2 a_3}^{\mu_1 \mu_2 \mu_3} \, = \, \sum_i f^{a_1 a_2}_{\phantom{a_1 a_2} e} \,
  f^{e a_3}_{\phantom{e a_3} b} \, {\bf T}_i^b \,
  \gamma_i^{\mu_1 \mu_2 \mu_3} (q_1, q_2; q_3) + (3 \leftrightarrow 1) + (3 \leftrightarrow 2) \ ,
\eeq 
with the kinematic function $\gamma_i(q_1,q_2;q_3)$ given by
\beq
\label{TripleCatKin}
  \gamma_i^{\mu_1\mu_2\mu_3} (q_1,q_2;q_3) & = & \frac{1}{\beta_i \cdot q_{123}} \,
  \Bigg\{ \frac{1}{12} \frac{\beta_i^{\mu_1} \beta_i^{\mu_2} \beta_i^{\mu_3}}{\beta_i \cdot q_2 \,
  \beta_i \cdot q_3 \, \beta_i \cdot q_{12}} \, \beta_i \cdot (3 q_3 - q_{12}) \nonumber \\
  & + & \frac{\beta_i^{\mu_3} \, \beta_i \cdot (q_3 - q_{12})}{\beta_i \cdot q_3 \beta_i \cdot q_{12}} \,  
  \frac{1}{q_{12}^2} \, \bigg( \frac{1}{2} \, g^{\mu_1 \mu_2} \beta_i \cdot q_1 + 
  \beta_i^{\mu_2} q_2^{\mu_1} \bigg) \nonumber \\
  & + & \frac{1}{q_{123}^2 \, q_{12}^2} \, \Big[ q_{12}^{2} \beta_i^{\mu_1} g^{\mu_2 \mu_3} + 
  2 q_2^{\mu_1} g^{\mu_2\mu_3} \beta_i \cdot (q_3 - q_{12}) + 
  4 q_3^{\mu_1} q_1^{\mu_2} \beta_i^{\mu_3} \nonumber \\
  & + & 4q_2^{\mu_1} \beta_i^{\mu_2} q_{12}^{\mu_3} + g^{\mu_1 \mu_2}
  \big( q_{23}^2 \beta_i^{\mu_3} + q_1^{\mu_3} \beta_i \cdot (q_{13} - 3 q_2) \big) \Big] \Bigg\} 
  - (1 \leftrightarrow 2) \, ,
\eeq 
where $q_{ij} \equiv q_i + q_j$, and $q_{123} \equiv q_1 + q_2 + q_3$. We note the inevitable complexity 
of this expression, which remains invariant under rescalings $\beta_i \to \kappa_i \beta_i$, as it must,
but otherwise depends on all possible Lorentz-invariants, $\beta_i \cdot q_j$ and $q_i \cdot q_j$, 
and carries different degrees of singularity in different collinear and soft limits, even if the overall
scaling is $\omega^{-3}$ for all terms.

In order to test the associativity of the OPE, we first translate this expression to celestial 
coordinates, after dotting it with gluon helicities, and then take two collinear limits in succession.


\subsection{Collinear limits for equal helicities}
\label{EqualHel}

We start with the all-positive helicities case, and we define
\beq
\label{Def3P}
  \gamma_{i}^{+++}(z_1, z_2; z_3) \, = \, 
  \vare^+_{\mu_1} (q_1) \, \vare^+_{\mu_2} (q_2) \, \vare^+_{\mu_3} (q_3) \,
  \gamma_i^{\mu_1 \mu_2 \mu_3}(z_1, z_2; z_3) \, .
\eeq
The result reads
\beq
\label{All3Plus}
  \gamma_{i}^{ +++}(z_1, z_2; z_3) & = & \frac{1}{\omega_1 |z_{1i}|^2 + 
  \omega_2 |z_{2i}|^2 + \omega_3 |z_{3i}|^2} \,
  \Bigg[ \frac{1}{12} \frac{\zbar_{1i} \zbar_{2i} \zbar_{3i} \, \big(3 \omega_3 |z_{3i}|^2 -
  \omega_1|z_{1i}|^2 - \omega_2|z_{2i}|^2 \big)}{\omega_2 \omega_3 |z_{2i}|^2 |z_{3i}|^2 
  \big(\omega_1 |z_{1i}|^2 + \omega_2 |z_{2i}|^2 \big)} 
  \nonumber \\
  && \hspace{5mm} + \, \frac{\zbar_{2i}\big(\omega_3 |z_{3i}|^2 - \omega_1 |z_{1i}|^2 - \omega_2|z_{2i}|^2 
  \big)}{2\omega_1 \omega_3 z_{12} z_{3i} \big(\omega_1 |z_{1i}|^2 + \omega_2 |z_{2i}|^2 \big)} 
  \nonumber \\
  && \hspace{5mm} + \, \frac{\omega_2 \zbar_{2i} \big(\omega_1 \zbar_{31} + \omega_2 \zbar_{32} \big) +
  \omega_1 \omega_3 \zbar_{31} \zbar_{3i}}{\omega_1 \omega_2 z_{12} \big( \omega_1 \omega_2 |z_{12}|^2 +
  \omega_1 \omega_3 |z_{13}|^2 + \omega_2 \omega_3 |z_{23}|^2 \big)} \Bigg] - 
  \big(1 \leftrightarrow 2 \big) \ .
\eeq
Note that the overall mass dimension of the $\gamma$ tensor is $(-3)$, so that the relevant quantity in
the celestial framework will be $\omega_1 \omega_2 \omega_3 \gamma_{i, +++}$. We now consider the collinear
limit $z_1 \rightarrow z_2$: retaining only singular contributions, we get
\beq 
\label{FirstColl3P}
    \gamma_{i}^{+++}(z_1, z_2; z_3) & \sim & 
    \frac{1}{z_{12}} \frac{1}{ \big( \omega_1 + \omega_2 \big) |z_{2i}|^2 + 
    \omega_3| z_{3i}|^2} \\
    && \times \, \Bigg[ \frac{\omega_3 |z_{3i}|^2 - \big( \omega_1 + \omega_2 \big) |z_{2i}|^2}{2\omega_1 
    \omega_3 z_{3i} \big( \omega_1 + \omega_2 \big) z_{2i}} +
    \frac{\omega_2 \zbar_{2i} \big( \omega_1 + \omega_2 \big) + \omega_1 \omega_3 \zbar_{3i}}{\omega_1
    \omega_2 \omega_3 \big( \omega_1 + \omega_2 \big) z_{32}} \Bigg] - (1 \leftrightarrow 2) \, .
    \nonumber
\eeq 
We then proceed to take the second collinear limit, which we chose to be $z_2 \rightarrow z_3$. Again retaining 
only singular contributions we find
\beq 
\label{SecColl3P}
  \gamma_{i}^{+++}(z_1, z_2; z_3) & \sim &
  - \, \frac{1}{\omega_1 \omega_2 \omega_3} \, \frac{1}{z_{12} \, z_{23} \, z_{3i}} \, .
\eeq 
Note that this has precisely the correct form to reconstruct the tree-level single soft current,
when plugged back into \eq{TripleCat1}. Indeed, recovering the color structure from \eq{TripleCat1},
multiplying times the three gluon energies, and reinterpreting the result from the perspective of 
celestial currents, we obtain a double OPE of the form
\beq 
\label{OPE3P}
    \bigg[ \Big[ j^{a_1} (z_1) \, j^{a_2} (z_2) \Big] \, j^{a_3} (z_3) \bigg]  \, \sim \, - \, 
    \frac{f^{a_1 a_2 e} f_e^{{\phantom e} a_3 b}}{z_{12} z_{23}} \, j_b (z_3) \, ,
\eeq 
where the square brackets denote the order in which collinear limits were taken. This structure is 
precisely what one would expect from the application in succession of two OPEs to a set of currents
generating a Ka\v c-Moody algebra: we have two single poles in the two collinear pairs, two structure 
constants with the appropriately linked indices, and a single (tree-level) current as the only operator 
left on the {\it r.h.s.}. If we now reverse the order of the collinear limits we easily find
\beq 
\label{AlOPE3P}
    \bigg[ j^{a_1} (z_1) \, \Big[ j^{a_2} (z_2) \, j^{a_3} (z_3) \Big] \bigg]  \, \sim \, - \, 
    \frac{f^{a_2 a_3 e} f_e^{{\phantom e} a_1 b}}{z_{23} z_{31}} \, j_b (z_1) \, ,
\eeq 
which confirms the associativity of the OPE in the case of three soft gluons with the same helicity. 
Naturally, an analogous result is obtained for the case of all negative helicities, upon swapping barred 
and non-barred variables and currents.


\subsection{Collinear limits for different helicities}
\label{DiffHel}

We now turn to the more subtle case of mixed helicities. In particular, we pick the assignments $(1,+)$, 
$(2, +)$, $(3,-)$, and we define
\beq
\label{Def2P}
  \gamma_{i}^{++-}(z_1, z_2; z_3) \, = \, 
  \vare^+_{\mu_1} (q_1) \, \vare^+_{\mu_2} (q_2) \, \vare^-_{\mu_3} (q_3) \,
  \gamma_i^{\mu_1 \mu_2 \mu_3}(z_1, z_2; z_3) \, .
\eeq
Taking the double limit in the first of the two orders considered above ({\it i.e.} first $z_1 \to z_2$
and then $z_2 \to z_3$) we get
\beq
\label{FirstColl2P}
  \gamma_{i}^{++-}(z_1, z_2; z_3) \, \sim \, - \, \frac{1}{z_{12}} \,
  \frac{1}{\omega_1 \omega_2 \omega_3 \big( \omega_1 + \omega_2 + \omega_3 \big)} \,
  \bigg[ \frac{\omega_3}{z_{23} \zbar_{3i}} + \frac{\omega_1 + \omega_2}{\zbar_{23} z_{3i}} \bigg] \, .
\eeq
This expression highlights the key difference arising upon introducing a different helicity: because 
the denominators in the brackets are different, there is no way of collecting terms in such a way as to 
cancel the sum of energies in the overall denominator, and we are forced to consider the result as the sum 
of two separate contributions. Introducing the energy fractions $y_i \equiv \omega_i/(\omega_1 + \omega_2 + 
\omega_3)$ (so that $\sum_i y_i = 1$) and multiplying times the product of the three energies we can 
write
\beq
\label{AlFirstColl2P}
  \omega_1 \omega_2 \omega_3 \, \gamma_{i}^{++-}(z_1, z_2; z_3) \, \sim \, - \, 
  \frac{1}{z_{12}} \,
  \bigg[ \frac{y_3}{z_{23} \zbar_{3i}} + \frac{y_1 + y_2}{\zbar_{23} z_{3i}} \bigg] \, .
\eeq
which suggest a celestial double OPE of the form
\beq 
\label{OPE2P}
  \bigg[ \Big[ j^{a_1} (z_1) \, j^{a_2} (z_2) \Big] \, \bar{j}^{a_3} (\bar{z}_3) \bigg]  
  \, \sim \, - \, 
  \frac{f^{a_1 a_2 e} f_e^{{\phantom e} a_3 b} }{z_{12}} \, \bigg[ \frac{y_1 + y_2}{\bar{z}_{23}} 
  \, j_b (z_2) + \frac{y_3}{z_{23}} \, \bar{j}_b (\bar{z}_3) \bigg] \, .
\eeq 
Notice that the factor inside the square bracket is exactly of the form of \eq{OPE+-}, as one can see by
using energy fractions also in that case, and thus re-writing $x_{12} = \omega_1/\omega_2 = y_1/y_2$, so 
that, for example, $1/(1 + x_{12}) = y_2$. Now however one of the two energies is replaced by the sum
$\omega_1 + \omega_2$, because the third current only sees the result of the OPE of the first two currents, 
which can be thought of as a single, effective, emission with energy $\omega_1 + \omega_2$. 
If we now take the two limits in reverse order, it is easy to verify that the result is
\beq 
\label{AlOPE2P}
  \bigg[ j^{a_1} (z_1) \, \Big[j^{a_2} (z_2) \, \bar{j}^{a_3} (\bar{z}_3) \Big] \bigg]  
  \, \sim \, - \, 
  \frac{f^{a_2 a_3 e} f_e^{{\phantom e} a_1 b}}{z_{31}} \, \bigg[ \frac{y_1}{\bar{z}_{23}} 
  \, j_b (z_2) + \frac{y_2 + y_3}{z_{23}} \, \bar{j}_b (\bar{z}_3) \bigg] \, .
\eeq 
The structure of \eq{OPE2P} and \eq{AlOPE2P} is the same but, crucially, the energy-dependent factors are 
different, so that the two limits do not commute. This simple calculation, entirely based on gauge-theory 
results, substantiates the findings of Ref.~\cite{Ball:2022bgg}: the OPE of the soft currents is not 
associative in the mixed-helicity sector of the theory. As in \secn{DoubleSoft}, this fact can be traced back 
to the non-abelian terms in \eq{TripleCat}: they force us to consider all possible orderings for the emission 
of the soft gluons and, in the case of gluons of different helicities, the cancellations in the kinematic 
dependence that rescue the same-helicity case cannot happen, because the two currents have different 
holomorphic character. Interestingly, the OPE problems discussed both in this Section and in \secn{DoubleSoft}
can be seen to originate from the simple but surprising structure of the suggested OPE connecting 
holomorphic and anti-holomorphic currents\footnote{The emergence of this structure was noticed also in 
Ref.~\cite{McLoughlin:2016uwa}.}. Setting aside for the moment the color structure, the OPEs are of the 
form
\beq
\label{GenProb}
  j(z) \, \bar{j} (\bar{w}) \, \sim \, \frac{\xi}{\bar{z} - \bar{w}} \, j (z) \, + \, 
    \frac{1 - \xi}{z - w} \, \bar{j} (z) \, ,
\eeq 
with $\xi$ a real parameter satisfying $0 \leq \xi \leq 1$, which in the present cases is related to the
energy fractions carried by different soft gluons. We note again that this would vanish in the case of
theories supporting holomorphic factorization. Conformal current algebras in which a similar structure 
seems to emerge were studied in Ref.~\cite{Ashok:2009xx}, and constitute a possible direction of further 
study.

In analogy to the case of the mixed-helicity double soft current, Eqs.~(\ref{OPE2P}) and (\ref{AlOPE2P})
are unambiguous but problematic, in this case because of the failure of associativity. Also here, one can
solve the problem of associativity by taking strongly-ordered limits, at the price of finding ambiguous 
results ({\it i.e.} results depending on the choice of strong ordering). Specifically, in order to recover 
associativity we can look at {\it completely ordered} limits, such as $\omega_1 \ll \omega_2 \ll \omega_3$, 
or $\omega_1 \gg \omega_2 \gg \omega_3$ (note that for $n>2$ soft particles one can also have {\it partially 
ordered} limits, such as $\omega_1 \sim \omega_2 \ll \omega_3$). For example, taking $\omega_1 \ll \omega_2
\ll\omega_3$, we find
\beq 
\label{OPE2PSO1}
  \bigg[ \Big[ j^{a_1} (z_1) \, j^{a_2} (z_2) \Big] \, \bar{j}^{a_3} (\bar{z}_3) \bigg]  
  \, \sim \, - \, 
  \frac{f^{a_1 a_2 e} f_e^{{\phantom e} a_3 b}}{z_{12} z_{23}} \, \bar{j}_b (\bar{z}_3) \, ,
\eeq 
and
\beq 
\label{AlOPE2PSO1}
  \bigg[ j^{a_1} (z_1) \, \Big[j^{a_2} (z_2) \, \bar{j}^{a_3} (\bar{z}_3) \Big] \bigg]  
  \, \sim \, - \, 
  \frac{f^{a_2 a_3 e} f_e^{{\phantom e} a_1 b}}{z_{23} z_{31}} \, \bar{j}_b (\bar{z}_3) \, .
\eeq 
On the other hand, taking $\omega_1 \gg \omega_2 \gg \omega_3$ we get
\beq 
\label{OPE2PSO2}
  \bigg[ \Big[ j^{a_1} (z_1) \, j^{a_2} (z_2) \Big] \, \bar{j}^{a_3} (\bar{z}_3) \bigg]  
  \, \sim \, - \, 
  \frac{f^{a_1 a_2 e} f_e^{{\phantom e} a_3 b}}{z_{12} \bar{z}_{23}} \, j_b (z_2) \, ,
\eeq 
and
\beq 
\label{AlOPE2PSO2}
  \bigg[ j^{a_1} (z_1) \, \Big[j^{a_2} (z_2) \, \bar{j}^{a_3} (\bar{z}_3) \Big] \bigg]  
  \, \sim \, - \, 
  \frac{f^{a_2 a_3 e} f_e^{{\phantom e} a_1 b}}{\bar{z}_{23} z_{31}} \, j_b (z_2) \, .
\eeq 
To summarize, the analysis of the triple soft current highlights new challenges for a celestial theory,
suggesting that the structure of the OPE for celestial currents should be both novel and intricate. This 
is not surprising from the standpoint of the bulk theory: the patterns of non-abelian radiation are known
to grow rapidly in complexity as the number of soft radiated particles grows. That notwithstanding, the 
structures emerging in collinear limits, and in strongly-ordered limits, even for triple radiation, are 
remarkably simple, and remarkably close to what would be expected in the context of a celestial theory.
Furthermore, the comparative simplicity of the expressions emerging in celestial coordinates, which in 
some cases is remarkable, suggests that the celestial framework could be a useful analysis tool also
for gauge-theory applications.


\section{Perspectives}
\label{Persp}

The celestial approach to the infrared structure of non-abelian gauge theories~\cite{Strominger:2013lka,
Strominger:2017zoo} has brought a novel and interesting viewpoint on an old but still very relevant and 
open set of problems. Remarkably, all soft divergences of non-abelian massless scattering amplitudes which 
have a color-dipole structure can be derived from a free-boson conformal field theory on the celestial sphere, 
to all orders in perturbation theory~\cite{Magnea:2021fvy}. Also remarkably, the tree-level soft current
for the emission of a single gluon in a general hard scattering process emerges from a conformal Ward 
identity in the same theory~\cite{Strominger:2013lka,Magnea:2021fvy}\footnote{We note that it was argued
in Ref.~\cite{Gonzo:2019fai} that using coherent states could, at least partially, account for loop 
corrections to the Ward identity.}. To what extent this approach generalizes to encompass further features 
of non-abelian soft physics remains an open question.

In order to tackle this question, we believe that it is crucial to make use of the vast set of perturbative
data that have accumulated over decades, concerning soft loop corrections to hard scattering processes,
as well as single and multiple soft radiation. Pointing to the highlights of this past work, the soft 
anomalous dimension matrix for massless scattering is fully known to three loops~\cite{Almelid:2015jia} 
(and was very recently calculated also for amplitudes with a massive leg in Refs.~\cite{Liu:2022elt, 
Gardi:2025lws}), the single soft gluon current is fully known to two loops~\cite{Dixon:2019lnw} and 
partially known to three loops~\cite{Herzog:2023sgb}, the double soft current is know to one 
loop~\cite{Zhu:2020ftr,Czakon:2022dwk}, and the triple soft emission current has been computed at tree 
level in~\cite{Catani:2019nqv}. These results pose strong constraints and serve as significant pointers 
for a possible celestial theory underpinning them.

In this paper, we focused our attention on soft currents, leaving a more detailed study of high-order virtual 
corrections to future work. We have collected, translated to celestial cooordinates, and analyzed, much of the 
available information on single and multiple soft radiation in massless non-abelian gauge theories. We emphasize
that our expectations on the celestial theory are limited to the soft sector, which, we believe, is naturally
interpreted holographically: thus, we expect a celestial theory to reproduce the color-correlated soft factor
of gauge amplitudes (as in Ref.~\cite{Magnea:2021fvy}), and to have predictive power for soft emission currents
(as in Ref.~\cite{Strominger:2013lka}), but we refrain from applying a celestial viewpoint to hard factors and 
hard radiation. Concretely, we consider directly the soft currents, regardless of the underlying hard process,
and we take further collinear or energy-ordered limits only on pre-determined soft factors: this is equivalent
to focusing on residues of celestial amplitudes at the leading-power singularity where the conformal weight
$\Delta \to 1$, as illustrated by the calculations in Ref.~\cite{Gonzalez:2020tpi}. This viewpoint has the 
further advantage that our results automatically apply to amplitudes with any number of external particles, 
with any spin, in any massless gauge theory, thanks to infrared universality and factorization. These features 
were shown to hold also for celestial amplitudes, in the conformally soft limit, in Ref.~\cite{Gonzalez:2021dxw}.

To summarize our main results, in \secn{SingSoft} we have considered the known loop corrections to the single
soft gluon current, whose tree-level expression follows from the Ward identity for the celestial free-boson
conformal theory. We found that one- and two-loop corrections are naturally represented in celestial coordinates,
in terms of which they take on simple and transparent expressions, given in \eq{CelCurr1lo2} and \eq{Dix2}. 
Crucially, we showed that, {\it to all orders}, all the logarithms that arise in the loop corrections to the 
single soft current, upon expansion in powers of $\e$, are UV logarithms, and they can all be reabsorbed in the 
renormalised running coupling, at the price of introducing a different physical scale for every color dipole.
From the standpoint of the bulk gauge theory, this is a well-understood and expected feature; from the point 
of view of the celestial theory, this results casts some doubt on the need to introduce a logarithmic CFT
on the sphere: one can argue, in fact, that physical scales should be considered as external to the CCFT, as 
was the case for the renormalization scale appearing in virtual corrections. On the other hand, a logarithmic 
CFT generating the relevant logarithms would be remarkable, since it would encode the running of the bulk gauge 
coupling. 

In \secn{DoubleSoft} we concentrated on the double soft current, at tree level and at one loop. Of special 
interest in this case are collinear limits for the two soft gluons, which map to the OPE of celestial 
currents, and energy-ordered limits, which provide significant simplifications. The results depend strongly on 
soft gluon polarizations: taking the collinear limit, the current for two same-helicity gluons suggests an 
OPE compatible with a Ka\v c-Moody symmetry (\eq{OPE++}), as was noticed already in Ref.~\cite{He:2015zea}.
In the case of opposite-helicity gluons the resulting celestial OPE (\eq{OPE+-}) is unambiguous, but very
unconventional: the holomorphic and the anti-holomorphic currents must have a non-vanishing OPE, with 
holomorphic and anti-holomorphic components weighted by factors depending on the energy fractions carried 
by the two soft gluons. Again, this is to be expected in the bulk gauge theory, where collinear splittings
are typically weighted by energy fractions: for example, considering the soft anomalous dimension matrix
for dipole correlations, the splitting anomalous dimension governing collinear limits displays energy-fraction
weights, that cannot be directly extracted from the free-boson CCFT (see, for example,~\cite{Becher:2009qa,
Magnea:2021fvy}). From the point of view of the celestial theory, \eq{OPE+-} resolves the ambiguity discussed
in Refs~\cite{He:2015zea,Freidel:2021ytz,Kapec:2022hih,Pranzetti:2025flv}, by providing a smooth interpolation
between the two possible strongly-ordered limits. This resolution however has a price, highlighted in 
\secn{TripleSoft}. When considering the one-loop corrections to the double  soft current, we note that 
the available information is still not sufficient to understand the best choice of physical scales for the 
running coupling, so that, in general, logarithms are still present. In the strongly-ordered limit, however,
the lengthy expression for the one-loop double current simplifies: upon further taking the collinear limit, 
the resulting expression (\eq{CollTot}) still respects the general structure of an OPE of currents, involving
both the tree-level and the one-loop single currents on the {\it r.h.s.}, and in this limit all logarithms 
can indeed be reabsorbed in the running coupling, which one can take as a positive sign. This structure 
is expected to generalize to higher perturbative orders.

In \secn{TripleSoft}, we explored the rather intricate structure of the triple emission current at tree level,
focusing on collinear limits for the most correlated, maximally non-abelian contributions. The resulting 
constraints on the OPE, based on the complete gauge-theory calculation of Ref.~\cite{Catani:2019nqv}, 
align with Ref.~\cite{Ball:2022bgg}: we find that same-helicity soft currents support the associativity of 
the OPE in double collinear limits, but this fails for currents involving opposite helicities. Interestingly, 
the mechanism for this failure is the same that emerged in the case of double currents: collinear limits
are unambiguous, but the resulting OPEs are weighted by energy fractions. Under double collinear limits, 
the dependence on energy fractions is different for different orders of limits. Associativity can be 
rescued by taking strongly-ordered limit, but that reinstates the ambiguity first noted in~\cite{He:2015zea}.

In balance, this study of soft currents shows that the celestial parametrization is both novel and insightful,
helping to display structures that are hard to uncover in the conventional QCD language: this was already 
apparent in~\cite{Magnea:2021fvy}. The evidence for the emergence of a simple structure at the celestial 
level, on the other hand, is mixed. The fact that most of the results we have displayed are broadly compatible 
with an OPE of celestial currents is, we believe, very non-trivial, and to some extent surprising. On the 
other hand, the OPE-like structures that emerge are highly unconventional, and seem to point to the need 
for substantial generalizations of the underlying framework, for example along the lines discussed 
in~\cite{Ball:2022bgg}. In this connection, we believe that the emerging OPE structure, as presented in 
\eq{GenProb}, should provide an important guideline in the search for the ultimate CCFT.

We also note that, on the gauge theory side, the simplifying features of the soft 
approximation are only associated with the emission of soft gluons from Wilson lines: once the soft radiation 
is emitted by a set of hard particles, its space-time evolution is not any longer governed by the soft 
approximation, but it is determined by the full dynamics of the gauge theory. In this sense, it would be 
truly remarkable if the highly correlated, dynamic evolution of soft radiation could be captured by a simple 
{\it product} of celestial currents. The evidence we have uncovered makes it --- in our view --- unlikely that 
a conventional CFT with just single-particle operators on the sphere could account in full generality for the complexity of the radiative results. Continuous distributions of color and energy on the celestial sphere, 
such as those emerging from observables like energy correlators (see, for example,~\cite{Jaarsma:2025tck}, 
and references therein), would seem like more natural candidates for celestial observables: their connections 
to soft gluon currents are however non-trivial. As an alternative, one could consider the possibility that the 
celestial theory only describes strongly ordered soft radiation, whose dynamics is significantly simpler 
(though still highly correlated). Ultimately, establishing the perimeter of applicability of the celestial 
theory will require more perturbative data, and possibly a more thorough re-analysis of existing data in 
celestial coordinates, as well as new insights.


\section*{Acknowledgements}

LM thanks Marco Meineri and Lorenzo Bianchi for useful discussions. LM was partially supported 
by the Italian Ministry of University and Research (MUR) through grant PRIN 2022BCXSW9.


\bibliographystyle{JHEP}
\bibliography{ConfIR2bib}


\end{document}